\documentclass{lmcs} 
\pdfoutput=1

\usepackage{lastpage}
\lmcsdoi{18}{4}{9}
\lmcsheading{}{\pageref{LastPage}}{}{}%
{Aug.~29,~2020}{Nov.~28,~2022}{}


\usepackage{amsmath,amsthm,amsfonts,amssymb,stmaryrd,bbm}
\usepackage[all,knot,poly]{xy}
\usepackage[utf8]{inputenc}

\def\ie{{\emph{i.e.}~}}

\definecolor{mred}{rgb}{0.7,0.1,0.1}
\definecolor{mblue}{rgb}{0,0,0.8}
\definecolor{mgreen}{rgb}{0,0.6,0.3}

\newcommand\Cr{\mathcal{C}}

\newcommand{\fl}{\rightarrow}
\newcommand{\dfl}{\Rightarrow}
\newcommand{\tfl}{\Rrightarrow}
\newcommand{\qfl}{\xymatrix@1@C=10pt{\ar@4 [r] &}}

\newcommand{\Ar}{\mathcal{A}}

\newcommand{\comp}{\star}
\newcommand{\ars}[1]{\rightarrow_{#1}}

\newcommand{\mult}{\odot}
\newcommand{\un}{{1}}
\newcommand{\uno}{\mathbbm{1}}
\newcommand{\lstar}{\overset{*}{\leftarrow}}
\newcommand{\rstar}{\overset{*}{\rightarrow}}
\newcommand{\rlstar}{\overset{*}{\leftrightarrow}}

\def\catego#1{{\bf{\sf #1}}}

\newcommand\tck[1]{{#1}^\top}
\newcommand\Sph{\mathrm{Sph}}

\newcommand{\dom}[0]{\mathit{d}}
\newcommand{\cod}[0]{\mathit{r}}
\newcommand{\adom}[0]{\mathit{ad}}
\newcommand{\acod}[0]{\mathit{ar}}
\renewcommand{\conv}[1]{{#1}^{\lor}}
\newcommand{\op}[1]{{#1}^{\mathit{op}}}
\newcommand{\con}[2]{{#1}^{\lor_{#2}}}

\newcommand{\fDia}[1]{| #1 \rangle}
\newcommand{\bDia}[1]{\langle #1 |}

\newcommand{\fBox}[1]{| #1 ]}
\newcommand{\bBox}[1]{[ #1 |}

\newcommand{\noeth}[1]{\mathcal{N}(#1)}
\newcommand{\noethi}[2]{\mathcal{N}_{#2}(#1)}

\def\lfl{\leftarrow}
\def\rfl{\rightarrow}

\newcommand{\rrs}[1]{#1^{c}}

\begin{document}

\title[Algebraic coherent confluence and higher globular Kleene algebras]{Algebraic coherent confluence\texorpdfstring{\\}{ }and higher globular Kleene algebras}

\author[C.~Calk]{Cameron Calk}[a]
\author[E.~Goubault]{Eric Goubault\lmcsorcid{0000-0002-3198-1863}}[a]
\author[P.~Malbos]{Philippe Malbos\lmcsorcid{0000-0003-4449-0091}}[b]
\author[G.~Struth]{Georg Struth\lmcsorcid{0000-0001-9466-7815}}[c,d]

\address{LIX, \'Ecole Polytechnique, CNRS, IP-Paris, Palaiseau, France}	
\email{ccalk@lix.polytechnique.fr, eric.goubault@polytechnique.edu}  

\address{Université Claude Bernard Lyon 1, CNRS UMR 5208, Institut Camille Jordan, 43 Blvd. du 11 novembre 1918, F-69622 Villeurbanne cedex, France}	
\email{malbos@math.univ-lyon1.fr}  

\address{Department of Computer Science, The University of Sheffield, Regent Court, 211 Portobello, Sheffield S1 4DP, UK}	
\email{g.struth@sheffield.ac.uk}  

\address{Collegium de Lyon, 26 Place Bellecour, 69002 Lyon, France}	
\email{g.struth@sheffield.ac.uk}  

\begin{abstract}
\noindent 
We extend the formalisation of confluence results in Kleene algebras
to a formalisation of coherent confluence proofs. For this, we
introduce the structure of higher globular Kleene algebra, a
higher-dimensional generalisation of modal and concurrent Kleene
algebra. We calculate a coherent Church-Rosser theorem and a coherent
Newman's lemma in higher Kleene algebras by equational reasoning. We
instantiate these results in the context of higher rewriting systems
modelled by polygraphs.
\end{abstract}

\maketitle

\section{Introduction}

Rewriting is a model of computation widely used in algebra, computer
science and logic. Rules of computation or algebraic laws are
described by \emph{rewrite relations} on symbolic or algebraic
expressions. Rewriting theory is strongly based on diagrammatic
intuitions. A central theme is the completion of certain
\emph{branching shapes} with \emph{confluence shapes} into
\emph{confluence diagrams}. Traditionally, the rewriting machinery has
been formalised in terms of algebras of binary relations: confluence
properties are described by union, composition and iteration
operations. A natural generalisation is given by Kleene algebras, in
which proofs of classical confluence results such as the Church-Rosser
theorem or Newman's lemma can be calculated~\cite{Struth02,Struth06,DesharnaisStruth04}. Beyond that, Kleene
algebras and similar structures are known for their ability to capture
complex computational properties by simple equational specifications
and reasoning~\cite{DoornbosBW97,Kozen97,Wright04,Struth08} and their
capacity to unify various semantics of computational interest,
including formal languages, binary relations, path algebras or
execution traces of automata~\cite{HofnerStruth10}.

Rewriting supports constructive proofs of coherence properties in
categorical algebra. In this setting, such properties are formulated
via a notion of contractibility for higher categories.  By contrast to
the standard diagrammatic and relational methods, coherence properties
can be generated by pasting a given set of higher-dimensional
witnesses for confluence or local confluence diagrams. The approach
has been initiated by Squier~\cite{Squier94} in the context of
homotopical finiteness conditions in string rewriting, and more recently
been extended to a method for higher or higher-dimensional
rewriting~\cite{GuiraudMalbos18}.  This method has been applied, for
instance, to give constructive proofs for coherence in
monoids~\cite{GaussentGuiraudMalbos15,HageMalbos17,HageMalbos22} and
for coherence theorems in monoidal
categories~\cite{GuiraudMalbos12mscs}.

Here we combine the two lines of research on Kleene-algebraic and
higher rewriting into a unified framework. We show how some
calculational confluence proofs in Kleene algebras, such as the
Church-Rosser theorem and  Newman's lemma, can be extended to
coherent confluence proofs.  To achieve this, we introduce higher
globular Kleene algebras with many compositions and domain and
codomain operations, which generalise both modal Kleene
algebras~\cite{DesharnaisStruth11} and concurrent Kleene
algebras~\cite{HoareMollerStruthWehrman11}.  These structures capture
the semantics of higher abstract rewriting algebraically.  We also
relate these generalised results to the point-wise approach of 
higher rewriting systems described by polygraphs. The main
contribution of this work is therefore the provision of a point-free
algebraic approach to coherence in higher rewriting that seems of
general interest in categorical algebra.

In this work, we only consider rewriting on \emph{strict} higher
categories, where all composition operations are strictly associative,
identities are strict under all compositions and all compositions
commute with each other, that is, all interchange laws are strict.
Calculating confluences in higher categories with a weakening of these
axioms remains a difficult open problem. It requires considering a
notion of higher rewriting modulo some axioms based on polygraphs
modulo certain relations~\cite{DupontMalbos22}, and the mechanisms of
rewriting modulo equations~\cite{Huet80}. Further, the higher globular
Kleene algebra structure that could be used for coherent confluence
proofs in a weak setting remains to be identified. We also point out
that this work does not address the decidability of equations in
higher categories.

\subsection*{Abstract coherent reduction}

Coherence proofs by rewriting are based on coherent formulations of
confluence results such as the Church-Rosser theorem and Newman's
lemma. We present the coherent extension of the former as an
example. An \emph{abstract rewriting system} on a set $X$ is, as
usual, a family ${\fl}=\{ \fl_i \}_{i\in I}$ of binary relations on
$X$.  It is \emph{confluent} if it satisfies the inclusion
\begin{equation}
\label{E:confluenceRelational}
\lstar \cdot \rstar \quad \subseteq \quad \rstar \cdot \lstar\;,
\end{equation}
where $\rstar$ denotes the reflexive, transitive closure of $\fl$, the
relation $\leftarrow$ denotes its converse and $\cdot$ denotes
relational composition. Moreover, $\fl$ has the \emph{Church-Rosser
  property} if the inclusion
\[
\rlstar \quad \subseteq \quad \rstar \cdot \lstar
\]
holds, where $\rlstar \ = ( \leftarrow \cup \rightarrow)^*$ denotes
the reflexive, symmetric, transitive closure of $\fl$. The
Church-Rosser theorem for $\fl$ states that these two inclusions
between relations are equivalent. It can be formulated more abstractly
in a Kleene algebra $K$ using the \emph{Kleene star} operation
$(-)^\ast:K\to K$, which generalises the reflexive, transitive closure
operation on relations~\cite{Struth02}, see
also~\eqref{SSS:ConfluenceKleeneAlgebras} below. Now, for all
$x,y\in K$,
\[
x^*\cdot y^* \leq y^* \cdot x^*
\quad\Leftrightarrow\quad
(x + y )^* \leq y^* \cdot x^*.
\]
The binary relations over a set $X$ form a Kleene algebra with respect
to relational composition, relational union, the reflexive transitive
closure operation, the empty relation and the unit relation.  The
Church-Rosser theorem for $\fl$ is thus an instance in the Kleene
algebra of binary relations for $x=\,\lfl$ and $y=\,\rfl$.

The diagrammatic interpretation of $\fl$ views an arrow $u\fl v$ as a
rewriting step whenever $(u,v)$ is an element of $\fl$. When $(u,v)$
is an element of $\rstar$ (resp. $\rlstar$), we say that $u$ is
related to $v$ by a \emph{rewriting sequence} (resp.  \emph{zig-zag
  sequence}) of finitely many rewriting steps.  We denote such
sequences by $f,g...$. A \emph{branching} (resp. \emph{confluence}) is
a pair $(f,g)$ (resp. $(f',g')$) of rewriting sequences of the shape
\[
\xymatrix@C=1.1em{
u_1
&
u
  \ar[l] _-{f}
  \ar[r] ^-{g}
&
v_1,
}
\qquad
\text{(resp. \quad}
\xymatrix@C=1.1em{
u_1
&
u'
  \ar@{<-}[l] _-{f'}
  \ar@{<-}[r] ^-{g'}
&
v_1
}
\text{\quad ).}
\]

The Church-Rosser theorem then states that, for all branchings $(f,g)$
of rewriting sequences, there exists a confluence $(f',g')$ if, and
only if, for any zig-zag sequence $h$ there exists a confluence
$(h',k')$:

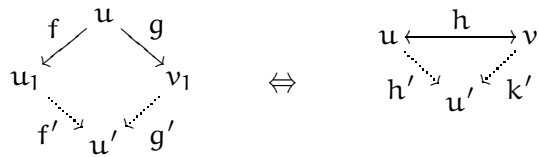
\[
\raisebox{1cm}{
\xymatrix@R=1.1em @C=1.1em{
&
u
\ar[dl] _-{f}
\ar[dr] ^-{g}
&
\\
u_1
\ar@{..>}[dr] _-{f'}
&
&
v_1
\ar@{..>}[dl] ^-{g'}
\\
&u'&
}}
\qquad\Leftrightarrow\qquad
\raisebox{0.7cm}{
\xymatrix@R=1.1em @C=1.1em{
u
\ar@{..>}[dr] _-{h'}
\ar@{<->}[rr] ^-{h}
&
&
v
\ar@{..>}[dl] ^-{k'}
\\
&
u'
&
}}
\]

By contrast to the relational Church-Rosser theorem, we can now no
longer use inclusions as witnesses of the forall/exist-relationships
between branchings or zig-zags and confluences in these
diagrams. Formally, we need to replace inclusions as $2$-cell in the
$2$-category $\catego{Rel}$ of relations by more general $2$-cells,
for which we write $\alpha,\beta...$. This leads to the coherent
Church-Rosser theorems of higher rewriting. In two dimensions, it
holds if there exists a set $\Gamma$ of $2$-dimensional cells such
that, if every branching can be completed to a confluence diagram
filled with elements of $\Gamma$ that are pasted together along their
$1$-dimensional borders, then every zig-zag sequence can be completed
to a Church-Rosser diagram filled with elements of $\Gamma$ that are
pasted along their $1$-dimensional borders (and of course vice
versa). Diagrammatically, for $2$-cells $\alpha$ and $\beta$ built
from $2$-cells in $\Gamma$,
\[
\raisebox{1cm}{
\xymatrix@R=1.25em @C=1.25em{
&
u
\ar[dl] _-{f}
\ar[dr] ^-{g}
&
\\
u_1
\ar@{..>}[dr] _-{f'}
&
&
v_1
\ar@{..>}[dl] ^-{g'}
\\
&u'&
\ar@2 "1,2"!<0pt,-20pt>;"3,2"!<0pt,+20pt> ^-{\alpha} 
}}
\qquad\Leftrightarrow\qquad
\raisebox{0.6cm}{
\xymatrix@R=2em @C=1.25em{
u
\ar@{..>}[dr] _-{h'}
\ar@{<->}[rr] ^-{h}
&
&
v
\ar@{..>}[dl] ^-{k'}
\\
&
u'
&
\ar@2 "1,2"!<+0pt,-7pt>;"2,2"!<0pt,10pt> ^-{\beta} 
}}
\]

\subsubsection*{Algebraic coherence}
The coherent Church-Rosser theorem constitutes one step in the proof
of Squier's theorem for higher rewriting systems, which provides a
constructive approach to \emph{coherence results} in higher
categories.  These are related to the fact that certain algebraic
properties of a categorical or algebraic structure may only hold up to
the existence of higher-dimensional morphisms. The classical coherence
conditions on associatiors and unitors in monoidal categories, for
example, require that if certain diagrams of natural isomorphisms
commute, then all the diagrams built from the corresponding natural
isomorphisms do.  A key issue is then the reduction of the property
``every diagram commutes'' to the property ``if a certain set of
diagrams each commute then every diagram
commute''~\cite{MacLane63,Stasheff63}.  For any collection of
higher-dimensional morphisms, coherence is thus the requirement that
the whole structure be contractible, that all parallel morphisms be
linked by higher morphisms. A coherence theorem states that, for each
generating collection of such morphisms, coherence is satisfied. An
objective is thus to obtain a minimal collection of generating higher
morphisms.

To solve coherence problems for monoids, formulated as two-dimensional
word problems, Squier introduced graph-theoretical methods on string
rewriting systems~\cite{Squier94}. His idea was to compute extensions
of string rewriting systems by homotopy generators, which model the
relations amongst rewriting sequences, so that every pair of zig-zag
sequences with same source and same target can be paved by composing
these generators. In Squier’s approach, the homotopy generators are
defined by the confluence diagrams of the critical branchings of the
string rewriting system, provided the string rewriting system is
convergent.
  
  \subsection*{Organisation and main results of the article}

\subsubsection*{Higher rewriting}

In Section~\ref{S:Preliminaries} we summarise notions from higher
rewriting. We first recall \emph{polygraphs}, which
represent  systems of generators and relations for higher categories
used for modelling higher coherence properties. Polygraphs, also
called \emph{computads}, were introduced by Street and
Burroni~\cite{Street76, DBLP:journals/tcs/Burroni93}. They are widely
used as rewriting systems that present higher algebraic
structures~\cite{Mimram14,GuiraudMalbos09}. Furthermore, polygraphs
allow formulating homotopical properties of rewriting systems through
polygraphic resolutions~\cite{Metayer03, GuiraudMalbos12advances}, as
well as coherence properties for
monoids~\cite{GuiraudMalbosMimram13,GaussentGuiraudMalbos15,GuiraudMalbos18},
higher categories~\cite{GuiraudMalbos09}, and monoidal
categories~\cite{GuiraudMalbos12mscs}. The latter are inspired by
Squier's approach to coherence results for monoids using convergent
string rewriting systems~\cite{Squier94}.

Formally, an \emph{$n$-polygraph} is a higher rewriting system made of
globular cells of dimension $0,1,\ldots , n$. It is defined recursively
as a sequence $P:=(P_0,P_1,\ldots,P_n)$, where for $0\leq k \leq n$,
the set $P_k$ consists of \emph{generating $k$-cells} of globular
shape:
\[
\xymatrix@C=4em{
s_{k-2}(\alpha)
	\ar@/^4ex/[r] ^-{s_{k-1}(\alpha)} _-{}="src"
	\ar@/_4ex/[r] _-{t_{k-1}(\alpha)} ^-{}="tgt"
&
t_{k-2}(\alpha)
\ar@2 "src"!<0pt,-15pt>;"tgt"!<0pt,+15pt> ^-{\alpha}
}
\]
The \emph{source} $s_{k-1}(\alpha)$ and \emph{target}
$t_{k-1}(\alpha)$ belong to the free $(k-1)$-category generated by the
underlying $(k-1)$-polygraph $(P_0,P_1,\ldots,P_{k-1})$.  A generating
$n$-cell $f : u \fl v$ in $P_n$ corresponds to an
\emph{$n$-dimensional rule}, reducing the $(n-1)$-cell $u$ to the
$(n-1)$-cell $v$.

The free category on the polygraph $P$, denoted by $P_n^\ast$, is the category of higher \emph{rewriting sequences} generated by the rules in $P_n$. Its $n$-cells are $(n-1)$-compositions
\[
f_1 \comp_{n-1} f_2 \comp_{n-1} \ldots \comp_{n-1} f_k
\]
of \emph{rewriting steps} with respect to $P_n$. The free
$(n,n-1)$-category on $P_n$, denoted by $\tck{P}_n$, is the category
of \emph{zig-zag sequences} generated by the rules in $P_n$, which
correspond to congruences between $(n-1)$-cells in $P_{n-1}^\ast$
modulo the rules in $P_n$.

In this work, we study the confluence properties of polygraphs by
considering \emph{cellular extensions} of the $n$-categories
$P_n^\ast$ and $\tck{P}_n$, whose elements are $(n+1)$-cells that are
\emph{confluence witnesses}. Formally, a cellular extension of the
free $n$-category $P_n^\ast$ (resp. free $(n,n-1)$-category
$\tck{P}_n$) consists of a set of globular $(n+1)$-cells that relate
the $n$-cells of $P_n^\ast$ (resp. $\tck{P}_n$).

\subsubsection*{Coherent confluence}
A \emph{branching} in an $n$-polygraph $P$, for $n\ge 1$,  is a pair $(f,g)$ of
$n$-cells of the free~$n$-category~$P_n^\ast$ with the same
$(n-1)$-source. A branching is \emph{local} when $f$ and $g$ are
rewriting steps.  A cellular extension $\Gamma$ of the free
$(n,n-1)$-category $P_n^\top$ is a \emph{confluence filler} of the
branching $(f,g)$ if there exist $n$-cells $f',g'$ in the free
$n$-category $P_n^\ast$ and two $(n+1)$-cells $\alpha$ and $\alpha'$

\[
\xymatrix@R=1.2em@C=1.2em{
&
u
\ar@{<-}[dl] _-{f^-}
\ar[dr] ^-{g}
&
\\
u_1
\ar[dr] _-{f'}
&
&
v_1
\ar@{<-}[dl] ^-{(g')^-}
\\
&u'&
\ar@2 "1,2"!<0pt,-20pt>;"3,2"!<0pt,+20pt> ^-{\alpha} 
}
\qquad\qquad
\xymatrix@R=1.2em@C=1.2em{
&
u
\ar[dl] _-{f}
&
\\
u_1
&
&
v_1
\ar[dl] ^-{g'}
\ar[ul] _-{g^-}
\\
&
u'
\ar[ul] ^-{(f')^-}
&
\ar@2 "1,2"!<0pt,-20pt>;"3,2"!<0pt,+20pt> ^-{\alpha'} 
}
\]
in the free $(n+1,n-1)$-category $\tck{P}_n[\Gamma]$ over $\tck{P}_n$
generated by $\Gamma$. The cellular extension $\Gamma$ is a
\emph{(local) confluence filler} for $P$ if it is a confluence filler
for each of its (local) branchings. Further, $\Gamma$ is a
\emph{confluence filler} of an $n$-cell $h$ in $P_n^\top$ if there
exist $n$-cells $h'$ and $k'$ in $P_n^\ast$ and an $(n+1)$-cell
$\alpha$ in the free $(n+1,n-1)$-category~$\tck{P}_n[\Gamma]$ of the
form \
\[
\xymatrix@R=1.5em{
u
\ar[dr] _-{h'}
\ar[rr] ^-{h}
&
&
v
\\
&
u'
\ar[ur] _-{(k')^-}
&
\ar@2 "1,2"!<+0pt,0pt>;"2,2"!<0pt,0pt> ^-{\alpha} 
}
\]
The cellular extension $\Gamma$ is a \emph{Church-Rosser filler} for
an $n$-polygraph $P$ if it is a confluence filler for every $n$-cell
in $P_n^\top$.

Theorem~\ref{T:CoherentCRARSFiller} below states that, for an
$n$-polygraph $P$, a cellular extension $\Gamma$ of $P_n^\top$ is a
confluence filler for $P$ if, and only if, $\Gamma$ is a Church-Rosser
filler for $P$.  Theorem~\ref{T:NewmanCoherentFiller} below states
that, when $P$ is \emph{terminating}, then $\Gamma$ is a local
confluence filler if, and only if, $\Gamma$ is a confluence filler
for~$P$. These statements are coherent, higher-dimensional extensions
of the Church-Rosser theorem and Newman's lemma, respectively.  In
Section~\ref{SS:GammaConfluenceAndFilling}, we relate these filler
properties to the standard coherent confluence properties used in
higher rewriting~\cite{GuiraudHoffbeckMalbos19}.

\subsubsection*{Modal and concurrent Kleeene algebras}

The forall/exist-relationships between higher-dimensional cells and
their sources and targets, expressed using various fillers, can be
captured algebraically through the \emph{higher globular Kleene
  algebras} introduced in Section~\ref{S:HigherDimKleene}.  Before
discussing them, we briefly review the modal Kleene
algebras~\cite{DesharnaisStruth11} and concurrent Kleene
algebras~\cite{HoareMollerStruthWehrman11} on which they are based.
  
Kleene algebras extend additively idempotent semirings
$(S,+,0,\cdot,1)$, in which addition models a notion of
nondeterministic choice or union and multiplication a non-commutative
composition, with a Kleene star $(-)^\ast$ that models a finite
repetition or iteration as a least fixpoint.  Models include binary
relations under union, relational composition and reflexive-transitive
closure, and sets of paths in a quiver or directed graph under union,
a complex product based on path composition and a Kleene star that
iteratively composes all paths in a given set with each other. Kleene
algebras allow specifying and proving the Church-Rosser theorem of
abstract rewriting~\cite{Struth06} using the fixpoint induction for
the Kleene star instead of the standard explicit induction on the
number of peaks in zig-zags.  Their path model forms the basis for
higher path algebras associated with polygraphic models of higher
rewriting.

Modal Kleene algebras equip Kleene algebras $K$ with forward and
backward modal operators introduced via domain and codomain operations
$\dom:K\to K$ and $\cod:K\to K$.  In the relational model, the domain
of a relation describes the set of all elements that it relates to
another element; its codomain describes those elements to which it
relates another element. In the path model, the domain of a set of
paths describes the set of all source elements of paths in the set,
and the codomain all target elements.  The relational model of Kleene
algebra provides the standard relational Kripke semantics of modal
diamond operators based on $\dom$ and $\cod$. The forward diamond
$\fDia{x}p =\dom(x\cdot p)$, for a relation $x$ and a set $p$, for
instance, models the set of all elements that may be related by $x$
with an element in $p$. In Kleene algebra, this generalises to
arbitrary elements $x$ and domain elements $p$, which are fixpoints of
the domain operator. Modal box operators, as duals of diamonds, can be
defined if the domain elements forms a Boolean algebra. They can be
based on antidomain and anticodomain operators, which model the
Boolean complements of domain and codomain operators. The antidomain
of a relation, for instance, models the set of elements that is does
not relate to any other element.  Noethericity and wellfoundnedness
can be expressed in modal Kleene algebras.  Newman's lemma for
abstract rewriting systems can therefore be proved in this
setting~\cite{DesharnaisStruth04}.

Finally, a \emph{concurrent Kleene
  algebra}~\cite{HoareMollerStruthWehrman11} is a double
Kleene algebra in which $+$ and $0$ are shared and the two
compositions $\cdot_0$ and $\cdot_1$ interact via a \emph{weak
  interchange law}
  \begin{equation*}
    (w\cdot_1 x)\cdot_0 (y\cdot_1 z) \le  (w\cdot_0 y)\cdot_1 (x\cdot_0 z),
  \end{equation*}
  and the two multiplicative units coincide. Typical models come from
  concurrency theory. They include shuffle language models from
  interleaving concurrency and partial-order-based models from
  non-interleaving concurrency.

\subsubsection*{Higher globular Kleeene algebras}

In Section~\ref{SS:ModalKleene}, we introduce a notion of globular
higher modal Kleene algebra. First, we define a \emph{$0$-dioid} as a
bounded distributive lattice, and for $n\geq 1$, an \emph{$n$-dioid}
as a family $(S,+,0,\mult_{i},\un_{i})_{0\leq i <n}$ of dioids, or
additively idempotent semirings, satisfying weak interchange laws
between the multiplications, akin to those of concurrent Kleene
algebras.  We then equip this structure with domain and codomain
operations $\dom_i, \cod_i : S \fl S$ for $0\leq i < n$, satisfying
typical axioms for $n$-categories such as $d_{i+1}\circ d_i = d_i$,
and $\cod_{i+1}\circ \cod_i=\cod_i$ for any $i$.

The domain and codomain operations yield forward and backward diamond
operators: for any $A \in S$, the $\fDia{A}_i, \bDia{A}_i$ are modal
operators on the \emph{$i$-dimensional domain algebra}
$S_i := \dom_i(S)$.  These are defined as usual and thus encode
higher-dimensional generalisations of the relational Kripke semantics:
$\fDia{A}_i \phi$, for instance, denotes the subset of $S_i$
containing the $i$-cells from which a set $A$ of $n$-cells may lead to
the set $\phi$ of $i$-cells. A concrete polygraphic model that
underpins these intuitions is introduced in
Section~\ref{SS:ModelsHigherMKA}.  In~\eqref{SSS:GlobularSemiring} we
impose conditions for globularity, conducing to the notion of
\emph{globular modal $n$-dioid}.

We further equip these structures with Kleene stars
$(-)^{\ast_i}: K \fl K$ for each $0\leq i < n$. These are \emph{lax
  morphisms} with respect to the $i$-multiplication of $j$-dimensional
elements on the right (resp.~left). Hence, for all $0 \leq i < j< n$,
all elements $A\in K$ and all $\phi \in K_j$ in the $j$-dimensional
domain algebra, 

\[
  \phi \mult_i A^{*_j} \leq (\phi \mult_i A )^{*_j} \qquad\text{ and
  }\qquad A^{*_j} \mult_i \phi \leq ( A \mult_i \phi )^{*_j}.
\]
The resulting structures are called \emph{globular modal $n$-Kleene
  algebras}.

In Section~\ref{SS:ModelsHigherMKA} we relate this structure to
polygraphs. We provide a model for higher Kleene algebras in the form
of a higher path algebra $K(P,\Gamma)$ induced by an $n$-polygraph $P$
and a cellular extension $\Gamma$.

\subsubsection*{Algebraic coherent confluence}

Section~\ref{S:AlgebraicCoherence} features our main results. After
revisiting the Church-Rosser theorem and Newman's lemma in modal
Kleene algebras in Section~\ref{SS:RewritingKleeneOne}, we define
notions of fillers in a globular modal $n$-Kleene algebra $K$
in~\eqref{SSS:ConfluenceFillers}. For $j$-dimensional elements
$\phi, \psi \in K_j := \dom_j(K)$, $A \in K$ is an
\emph{$i$-confluence filler} (resp.~\emph{$i$-Church-Rosser filler})
for $(\phi, \psi)$ if
\[
\fDia{A}_j(\psi^{*_{i}} \mult_{i} \phi^{*_{i}}) \geq  \phi^{*_{i}} \mult_{i} \psi^{*_{i}} \qquad 
(\text{resp. } \fDia{A}_j(\psi^{*_{i}} \mult_{i} \phi^{*_{i}}) \geq (\psi + \phi)^{*_{i}}).
\]
The property on the left states that the set of all $i$-cells for
which there exists an $i$-confluence for $(\phi,\psi)$ with witness
$A$ contains the $i$-branching for $(\phi,\psi)$. The explanation for
the property on the right is analogous. We define a notion of
\emph{local $i$-confluence filler} along the same lines.

We introduce a notion of whiskering in $n$-Kleene algebras
in~\eqref{SSS:Whiskers}. We define, for $\phi, \psi \in K_j$ and an
$i$-confluence filler $A \in K$ of $(\phi, \psi)$, the
\emph{$j$-dimensional $i$-whiskering} of $A$ as
\begin{equation*}
\hat{A} := (\phi+ \psi)^{*_i}\mult_i A \mult_i (\phi+ \psi)^{*_i}.
\end{equation*}

We then prove two variants of the coherent Church-Rosser theorem in
globular $n$-Kleene algebras.  The first, Proposition~\ref{Th:CRInd}, uses an
explicit inductive argument external to the $n$-Kleene structure,
based on powers that can be defined in any $n$-semiring. For
$0\leq i < j < n$, it states that for every $\phi, \psi \in K_j$,
every $i$-confluence filler $A$ of $(\phi, \psi)$ and every natural
number $k$ there exists an $A_k \leq \hat{A}^{*_j}$ such that
\[
\cod_j(A_k) \leq   \psi^{*_i} \phi^{*_i}
\qquad\text{and}\qquad
\dom_j(A_k) \geq  (\phi + \psi)^{k_i},
\]
where $(\phi + \psi)^{0_i} = \un_i$ and $(\phi + \psi)^{k_i} = (\phi + \psi) \mult_i (\phi + \psi)^{k_i -1}$.

By constrast, the proof of the second theorem relies only on the
internal fixpoint induction given by the axioms for the Kleene star.
It constitutes our first main result.

\begin{quote}
  \textbf{Theorem~\ref{Th:CR}.}  \emph{ Let $K$ be a globular
    $n$-modal Kleene algebra and $0\leq i<j<n$. Then, for every
    $\phi, \psi \in K_j$ and every $i$-confluence filler $A\in K$ of
    $(\phi,\psi)$,
\[
\fDia{\hat{A}^{*_j}}_j (\psi^{*_i}\phi^{*_i}) \geq (\phi + \psi)^{*_i}.
\]
Thus $\hat{A}^{*_j}$ is an $i$-Church-Rosser filler for $(\phi,\psi)$.
}
\end{quote}

In Section~\ref{SS:CoherentNewman}, we introduce notions of
termination and well-foundedness in $n$-Kleene algebras in which the
domain algebras $K_i$ have a Boolean structure for all $i \leq p <
n$. This leads to our second main result: a specification and proof of
a coherent Newman's lemma in such algebras.

\begin{quote}
\textbf{Theorem \ref{Th:Newman}.}
\emph{
Let $0 \leq i \leq p < j < n$, and let $K$ be a globular $p$-Boolean modal Kleene algebra such that
\begin{enumerate}
\item $(K_i, +, 0, \mult_i, \un_i, \neg_i)$ is a complete Boolean algebra,
\item $K_j$ is continuous with respect to $i$-restriction, that is,
  for all $\psi,\psi'\in K_j$ and every family
  $(p_\alpha)_{\alpha\in I}$ of elements of $K_i$ such that
  $sup_I (p_\alpha)$ exists, 
\[
\psi \mult_i sup_I (p_\alpha) \mult_i \psi' 
= sup_I (\psi \mult_i p_\alpha \mult_i \psi').
\]
\end{enumerate}
Then, for any $\psi\in K_j$ $i$-Noetherian, and $\phi \in K_j$
$i$-well-founded, if $A$ is a local $i$-confluence filler for
$(\phi, \psi)$, then
\[
\fDia{\hat{A}^{*_i}}_j(\psi^{*_i}\phi^{*_i}) \geq \phi^{*_i}\psi^{*_i}.
\]
Thus $\hat{A}^{*_j}$ is an $i$-confluence filler for $(\phi,\psi)$.
}
\end{quote}

Finally, in Section~\ref{SS:ApplicationRewriting}, we instantiate
these results in the context of higher abstract rewriting, using the
higher-dimensional path model defined in
Section~\ref{SS:ModelsHigherMKA}.

\subsection*{Outlook}

\subsubsection*{Toward an algebraic Squier's theorem}

Our results provide formal equational proofs of the coherent
Church-Rosser theorem and the coherent Newman's lemma in higher
globular Kleene algebras. These are the main ingredients in the proof
of Squier's coherence theorem~\cite{Squier94} for string rewriting
systems, used in constructive proofs of coherence in categorical
algebra.  It remains to formalise this result within the higher Kleene
algebras framework.  A first obstacle is the formalisation of the
coherent critical branching lemma, stating that local coherent
confluence is equivalent to coherence confluence of all critical
branchings. This requires taking the algebraic and syntactic nature of
terms in the rewriting system into account~\cite{Nivat73, BookOtto93,chenavier_dupont_malbos_2021}. 
This remains an open
problem in formalisms such as Kleene algebras. In particular, it would
be interesting to identify the enrichment of the Kleene algebra
structure needed for formalising the critical confluence property of
string or term rewriting systems.

\subsubsection*{Formalisation of cofibrant replacements}

A second obstacle is to capture normalisation strategies in higher
Kleene algebras algebraically~\cite{CalkGoubaultMalbos21}.  Squier's
coherence theorem is the first step in the construction of cofibrant
replacements of algebraic structures using convergent
presentations~\cite{GuiraudMalbos12advances}.  We expect that the
material introduced in this article will enable us to give an
algebraic formalisation of acyclicity, which could in turn yield an
algebraic criterion for cofibrance.

\subsubsection*{Formalisation of cofibrant replacements in proof assistants} 
The results of this article are part of a research program that aims
at developing constructive methods for higher algebras based on
rewriting. The aim is to formalise, by rewriting, the computation in
internal monoids of monoidal categories, which categorify the
associative rewriting paradigm. This framework generalises word and
term rewriting, linear rewriting, operadic and propadic rewriting. The
overall goal is to compute cofibrant replacements of these structures
by rewriting and to formalise these computations. In this article, we
formalise the abstract coherent Church-Rosser and Newman theorems in
globular Kleene algebras. The Knuth-Bendix procedure provides a
characterisation of local confluence for algebraic rewriting systems
in terms of critical branchings. Our aim is to extend the
formalisation of the abstract case to coherent rewriting systems of
internal monoids. We expect to implement the proofs of the coherence
theorems in higher rewriting with Isabelle, Coq or Lean.

Another objective an algebraic formalisation of normalisation
strategies in rewriting. These allow building cofibrant
$(\omega,1)$-categorical replacements of algebraic structures
presented by confluent and terminating rewriting
systems~\cite{GuiraudMalbos12advances}. We expect that these
constructions can be formalised in $\omega$-globular Kleene algebras.

\section{Preliminaries on higher rewriting}
\label{S:Preliminaries}

In this preliminary section, we introduce the relevant notions of
higher rewriting. In its two subsections we recall the definition of
polygraphs and their properties as rewriting systems presenting higher
categories. In Section~\ref{SS:CoherenceConfluence} we introduce
the notion of \emph{confluence filler} for polygraphs with respect to
cellular extensions. We then formulate and give point-wise proofs of
the coherent versions of the Church-Rosser theorem and Neman's lemma
in the polygraphic setting. Finally, in its last subsection, we relate
the confluence filler property to a more standard coherent confluence
property~\cite{GuiraudHoffbeckMalbos19}.

\subsection{Polygraphs}

We first recall basic notions of polygraphs~\cite{Burroni93}, also
called \emph{computads} in~\cite{Street76}, see
also~\cite{Metayer03,GuiraudMalbos12advances}. Yet we start from
higher categories and refer to standard textbooks for
details~\cite{Leinster04,MacLane98}.

\subsubsection{Higher categories}
\label{SSS:HigherCategories}
Let $n$ be a natural number.  A \emph{(strict globular) $n$-category}
$\Cr$ consists of the following data.
\begin{enumerate}
\item It is a reflexive $n$-globular set, that is, a diagram of sets
  and functions of the form
\[
\xymatrix @C=3.3em @W=2.5em{
\Cr_0 
	\ar [r] |-*+{\iota_1}
&
\Cr_1 
	\ar@<-1.3ex> [l] _-{s_0}
	\ar@<1.3ex> [l] ^-{t_0}
	\ar [r] |-*+{\iota_2}
& 
\quad
\cdots
\quad
	\ar@<-1.3ex> [l] _-{s_1}
	\ar@<1.3ex> [l] ^-{t_1}
	\ar [r] |-*+{\iota_{n-1}}
& 
\;\;\Cr_{n-1}\;\;
	\ar@<-1.3ex> [l] _-{s_{n-2}}
	\ar@<1.3ex> [l] ^-{t_{n-2}}
	\ar [r] |-*+{\iota_n}
&
\Cr_n
	\ar@<-1.3ex> [l] _-{s_{n-1}}
	\ar@<1.3ex> [l] ^-{t_{n-1}}
}
\]
whose functions $s_i, t_i: \Cr_{i+1} \fl \Cr_i$ and $\iota_i : \Cr_{i-1} \fl \Cr_{i}$ satisfy the \emph{globular relations}
\begin{equation}
\label{E:GlobularLaw1}
s_i\circ s_{i+1} = s_i \circ t_{i+1},
\qquad
t_i\circ s_{i+1} =t_i \circ t_{i+1}
\end{equation}
and the \emph{identity relations}
\begin{equation}
\label{E:GlobularLaw2}
s_i\circ\iota_{i+1} = id_{\Cr_i},
\quad
t_i\circ\iota_{i+1} = id_{\Cr_i}.
\end{equation}
\item It is equipped with the structure of a category on
\[
\xymatrix @C=3em @W=2.5em{
\Cr_k
&
\Cr_\ell
	\ar@<-0.7ex> [l] _-{s_k^\ell}
	\ar@<0.7ex> [l] ^-{t_k^\ell}
}
\]
for all $k<\ell$, where 
\[
s_k^\ell := s_k\circ \ldots \circ s_{\ell-2}\circ s_{\ell-1}
\qquad\text{ and }\qquad
t_k^\ell := t_k\circ \ldots \circ t_{\ell-2} \circ t_{\ell-1},
\]
and whose \emph{$k$-composition morphism} on $\Cr_\ell$ is denoted by $\star_k^\ell: \Cr_\ell\star_k \Cr_\ell \fl \Cr_\ell$.
\item The $2$-globular set
\[
\xymatrix @C=3em @W=2.5em{
\Cr_j
&
\Cr_k 
	\ar@<-0.7ex> [l] _-{s_j^k}
	\ar@<0.7ex> [l] ^-{t_j^k}
& 
\Cr_\ell
	\ar@<-0.7ex> [l] _-{s_k^\ell}
	\ar@<0.7ex> [l] ^-{t_k^\ell}
}
\]
is a $2$-category for all $j<k<\ell$, see~{\cite[XII. 3.]{MacLane98}}. 
\end{enumerate}

\subsubsection{Notations}
\label{SSS:Notations}

The elements of $\Cr_k$ are called \emph{$k$-cells of $\Cr$}.  For
$0\leq k < n$, we abuse notation, denoting by $\Cr_k$ the underlying
$k$-category of $k$-cells of $\Cr$. The maps $s_i, t_i$ and $\iota_i$
are called \emph{source}, \emph{target} and \emph{unit} maps
respectively.  For a $k$-cell $f$ of $\Cr$ and for $0\leq i < k$, we
call $s_i(f)$ (resp. $t_i(f)$) the \emph{$i$-source}
(resp. \emph{$i$-target}) of $f$. We denote the identity $(k+1)$-cell
of $\iota_{k+1}(f)$ by $1_f$.  When $f$ and $g$ are
\emph{$i$-composable} $k$-cells, for $i<k$, that is when
$t_i(f)=s_i(g)$, we denote their $i$-composite by $f\star_i g$.  By
condition {\bf iii)}, the compositions satisfy the \emph{interchange
  law}
\begin{equation}
\label{E:ExchangeLaw}
(f\star_j f') \star_k (g\star_j g') = (f\star_k g) \star_j (f'\star_k g'),
\end{equation}
for all $0\leq j < k <n$, and whenever all compositions are defined.

The $(k-1)$-composition of $k$-cells $f$ and $g$ is denoted by
juxtaposition $fg$, and the $(k-1)$-source $s_{k-1}(f)$ and the
$(k-1)$-target $t_{k-1}(f)$ of a $k$-cell $f$ are denoted by $s(f)$
and $t(f)$, respectively.  To highlight the relative dimensions of
cells, we denote cells by single arrows $\fl$, double arrows $\dfl$,
and triple arrows $\tfl$. In particular, if we denote a $k$-cell in
$\Cr$ by $f : u\dfl v$, then we denote $(k-1)$-cells of $\Cr$ by
$u : p \fl q$ the and the $(k+1)$-cells of $\Cr$ by $A: f \tfl g$ in
to distinguish their dimensions notationally. Such globular cells are
depicted as follows:
\[
\xymatrix @C=3.5em @!C{
p
   \ar @/^5ex/ [rr] ^{u} _{}="src1"
   \ar @/_5ex/ [rr] _{v} _{}="tgt1"
&&
q
\ar@2 "src1"!<-20pt,-15pt>;"tgt1"!<-20pt,15pt> _{f} ^{}="srcA"
\ar@2 "src1"!<20pt,-15pt>;"tgt1"!<20pt,15pt> ^{g} _{}="tgtA"
\ar@3 "srcA"!<10pt,0pt>;"tgtA"!<-10pt,0pt> ^*+{A}
}
\]
The globular relations~\eqref{E:GlobularLaw1} imply that any $k$-cell
$f$ has globular shape:
\[
\xymatrix @C=0.8em{
s_i\circ s_{i+1}(f) = s_i \circ t_{i+1}(f)
   \ar @/^5ex/ [rr] ^{s_{i+1}(f)} _{}="src1"
   \ar @/_5ex/ [rr] _{t_{i+1}(f)} _{}="tgt1"
&&
t_i\circ s_{i+1}(f) =t_i \circ t_{i+1}(f)
\ar@2 "src1"!<0pt,-10pt>;"tgt1"!<0pt,+10pt> _{f} ^{}="srcA"
}
\]
With this diagrammatic notation, the interchange law~\eqref{E:ExchangeLaw}, for instance, becomes
\[
\xymatrix @C=4em{
p
  \ar@/^4ex/ [r] _{}="src1"
  \ar[r] ^{}="tgt1" _{}="src2"
  \ar@/_4ex/ [r] ^{}="tgt2"
&
q
\ar@2 "src1"!<0pt,0pt>;"tgt1"!<0pt,0pt> ^-{f}
\ar@2 "src2"!<0pt,0pt>;"tgt2"!<0pt,0pt> ^-{f'}
}
\;\star_k\;
\xymatrix @C=4em{
q
  \ar@/^4ex/ [r] _{}="src1"
  \ar[r] ^{}="tgt1" _{}="src2"
  \ar@/_4ex/ [r] ^{}="tgt2"
&
r
\ar@2 "src1"!<0pt,0pt>;"tgt1"!<0pt,0pt> ^-{g}
\ar@2 "src2"!<0pt,0pt>;"tgt2"!<0pt,0pt> ^-{g'}
}
\qquad
=
\qquad
\begin{tabular}{c}
\xymatrix @C=4em{
p
  \ar@/^2ex/ [r] _{}="src1"
  \ar@/_2ex/ [r] ^{}="tgt1"
  &
q
  \ar@/^2ex/ [r] _{}="src2"
  \ar@/_2ex/ [r] ^{}="tgt2"
  &
r
\ar@2 "src1"!<0pt,0pt>;"tgt1"!<0pt,0pt> ^-{f}
\ar@2 "src2"!<0pt,0pt>;"tgt2"!<0pt,0pt> ^-{g}
}
\\
$\star_j$
\\
\xymatrix @C=4em{
p
  \ar@/^2ex/ [r] _{}="src1"
  \ar@/_2ex/ [r] ^{}="tgt1"
  &
q
  \ar@/^2ex/ [r] _{}="src2"
  \ar@/_2ex/ [r] ^{}="tgt2"
  &
r
\ar@2 "src1"!<0pt,0pt>;"tgt1"!<0pt,0pt> ^-{f'}
\ar@2 "src2"!<0pt,0pt>;"tgt2"!<0pt,0pt> ^-{g'}
}

\end{tabular}
\]

\subsubsection{Identities and whiskers}
\label{SSS:IdentitiesWhiskerings}
Given a $k$-cell $f$, the identity $l$-cell on $f$ for
$k\leq l \leq n$ is denoted by~$\iota^l_k(f)$ and defined by
induction, setting $\iota^{k}_k(f) := f$ and
$\iota^l_k(f) := 1_{\iota_k^{l-1}(f)}$ for $k < l \leq n$. In this way,
for $0\leq k < l \leq n$, we associate a unique identity cell
$\iota_k^l(f)$ of dimension $l$ to every $k$-cell $f$, which is called
the \emph{$l$-dimensional identity} on $f$.

In higher categories, such iterated identities are important
for defining compositions.  For $0\leq i< k < l \leq n$, a $k$-cell
$f$ and a $l$-cell $g$ such that $t_i(f)=s_i(g)$, the $i$-composite of
$f$ and $g$ is defined as
\[
f\star_i g = \iota_k^l(f) \star_i g.
\]
If $t_i(g) = s_i(f)$, we define $g\star_i f = g \star_i \iota_k^l(f)$.

For $0\leq i < j \leq k$, an \emph{$(i,j)$-whiskering} of a $k$-cell
$f$ is a $k$-cell $\iota_j^k(u) \star_i f \star_i \iota_j^k(v)$, where
$u$ and $v$ are $j$-cells, as in the diagram
\[
\xymatrix @C=2.4em{
s_i(u)
\ar[r] ^-{u} _{}="src0"
&
s_{j-1}(f)
   \ar @/^5ex/ [rr] ^{s_{j}(f)} _{}="src1"
   \ar @/_5ex/ [rr] _{t_{j}(f)} _{}="tgt1"
&&
t_{j-1}(f)
\ar[r] ^-{v} _{}="src2"
&
t_i(v)
\ar@2 "src1"!<0pt,-10pt>;"tgt1"!<0pt,+10pt> _{f}
\ar@2@(dl,dr) "src0"!<0pt,0pt>;"src0"!<0pt,0pt> _{\iota_j^k(u)}
\ar@2@(dl,dr) "src2"!<0pt,0pt>;"src2"!<0pt,0pt> _{\iota_j^k(v)}
}
\]
To simplify notation, we denote this $k$-cell by $u \star_i f \star_i v$.  A
$(k-1,k-1)$-whiskering $1_u\star_{k-1} f \star_{k-1} 1_v$ of a
$k$-cell $f$ is called a \emph{whiskering} of $f$ and denoted by
$ufv$.

\subsubsection{$(n,p)$-categories}
\label{SSS:(n,p)-categories}
If $\Cr$ is an $n$-category and $0\leq i < k \leq n$, a $k$-cell $f$
of~$\Cr$ is \emph{$i$-invertible} if there exists a $k$-cell $g$
in $\Cr$ with $i$-source $t_i(f)$ and $i$-target $s_i(f)$ in $\Cr$
called the \emph{$i$-inverse of $f$}, which satisfies

\[
f\star_i g \:=\: 1_{s_i(f)}
\qquad\text{and}\qquad
g\star_i f \:=\: 1_{t_i(f)}.
\]
The $i$-inverse of a $k$-cell is necessarily unique.  When $i=k-1$, we
say that $f:u\fl v$ is \emph{invertible} and we denote its
$(k-1)-$inverse by $f^{-1}:v\fl u$ or $f^-:v\fl u$ for short, which we simply call its \emph{inverse}.
If in addition the $(k-1)$-cells $u$ and $v$ are invertible, then there
exist $k$-cells
\[
u^-\star_{k-2} f^- \star_{k-2} v^- : u^- \fl v^-,
\qquad
v^-\star_{k-2} f^- \star_{k-2} v^- : u^- \fl v^-
\]
in $\Cr$.
For a natural number $p\leq n$, or for $p=n=\infty$, an
\emph{$(n,p)$-category} is an $n$-category whose $k$-cells are
invertible for every $k>p$. When $n<\infty$, this is a $p$-category
enriched in $(n-p)$-groupoids and, when $n=\infty$, a $p$-category
enriched in $\infty$-groupoids. 

\subsubsection{Spheres and cellular extensions}
Let $\Cr$ be an $n$-category.
A \emph{$0$-sphere} of $\Cr$ is a pair of $0$-cells of $\Cr$. For
$1\leq k \leq n$, a \emph{$k$-sphere} of $\Cr$ is a pair $(f,g)$ of
$k$-cells such that $s_{k-1}(f)=s_{k-1}(g)$ and
$t_{k-1}(f)=t_{k-1}(g)$. We denote by $\Sph_k(\Cr)$ the set of
$k$-spheres of $\Cr$.

A \emph{cellular extension} of $\Cr$ is a set $\Gamma$ equipped with a
map $\partial : \Gamma \fl \Sph_n(\Cr)$. For $\alpha \in \Gamma$, the
\emph{boundary} of the sphere $\partial(\alpha)$ is denoted
$(s_n(\alpha),t_n(\alpha))$, defining in this way two maps
$s_n,t_n : \Gamma \fl \Cr_n$ satisfying the globular relations
\[
s_{n-1}\circ s_n = s_{n-1} \circ t_n \qquad \text{and} \qquad t_{n-1}\circ s_n = t_{n-1} \circ t_n.
\] 

The free \emph{$(n+1)$-category over $\Cr$ generated by the cellular extension $\Gamma$} is
the $(n+1)$-category, denoted by $\Cr[\Gamma]$ and defined as follows:
\begin{enumerate}[{\bf i)}]
\item its underlying $n$-category is $\Cr$,
\item its $(n+1)$-cells are built as formal $i$-compositions, for $0\leq i \leq n$, of elements of $\Gamma$ and $k$-cells of $\Cr$, seen as $(n+1)$-cells with source and target in
$\Cr_n$.  
\end{enumerate}
The \emph{quotient} of the $n$-category $\Cr$ by $\Gamma$, denoted by $\Cr/\Gamma$, is the $n$-category we obtain from $\Cr$ by identifying the $n$-cells $s_n(\alpha)$ and $t_n(\alpha)$, for every $n$-sphere $\alpha$ of $\Gamma$.

The free \emph{$(n+1,n)$-category over $\Cr$ generated by $\Gamma$}, denoted by $\Cr(\Gamma)$, is defined by 
\[
\Cr(\Gamma) = \Cr[\Gamma,\Gamma^-]/\mathrm{Inv}(\Gamma),
\]
where 
\begin{enumerate}[{\bf i)}]
\item $\Gamma^-$ is the cellular extension of $\Cr$ made of spheres $\alpha^- =(t_n(\alpha),s_n(\alpha))$, for each $\alpha$ in $\Gamma$, 
\item $\mathrm{Inv}(\Gamma)$ is the cellular extension of the free $(n+1)$-category $\Cr[\Gamma,\Gamma^-]$, made of $(n+1)$-spheres
\[
(\alpha \star_{n} \alpha^- , 1_{s_n(\alpha)}),
\qquad
(\alpha^-\star_{n} \alpha, 1_{t_n(\alpha)}).
\]
\end{enumerate}
  
We refer to~\cite{Metayer03} for explicit free constructions on
cellular extensions over $n$-categories.

\subsubsection{$n$-polygraphs}
Polygraphs are models of free higher categories. They are defined by induction on the dimension.
For $n\geq 0$, an \emph{$n$-polygraph} $P$ consists of a set $P_0$ and for every $0\leq k <n$ a cellular extension $P_{k+1}$ of the \emph{free $k$-category} 
\[
P_0[P_1]\ldots [P_k].
\]
For $0\leq k \leq n$, the elements of $P_k$ are called the \emph{generating $k$-cells} of $P$. 

The \emph{free $n$-category} $P_0[P_1]\ldots [P_{n-1}][P_n]$ (\emph{resp.} the \emph{free $(n,n-1)$-category} $P_0[P_1]\ldots \linebreak {[P_{n-1}]}(P_n)$) \emph{generated by $P$} will be denoted by $P_n^*$ (\emph{resp.} $P_n^\top$).
We refer to~\cite{Metayer03} for the details of the free constructions
on an $n$-polygraph. Note that a $0$-polygraph is a set and
an $1$-polygraph corresponds to a directed graph, whose set of vertices
is $P_0$ and $P_1$ is the set of arrows $f$ with source $s_0(f)$ and
target $t_0(f)$.

\subsection{Rewriting properties of polygraphs}

\subsubsection{Polygraphic rewriting}
\label{SSS:PolygraphicRewriting}
A \emph{rewriting step} of an $n$-polygraph $P$ is an $n$-cell of the free $n$-category~$P_n^\ast$ of the form
\[
u_{n-1}\comp_{n-2}(u_{n-2}\comp_{n-3}\ldots \comp_2 (u_2\comp_1(u_1\comp_0 f \comp_0 v_1)\comp_1 v_2) \comp_2 \ldots \comp_{n-3} v_{n-2})\comp_{n-2} v_{n-1},
\]
for a generating $n$-cell $f$ in $P_n$ and $i$-cells $u_i,v_i$ in $P_n^\ast$, with $1\leq i <n$. We denote by $\rrs{P_n}$ the set of rewriting steps of $P$. An $(n-1)$-cell $u$ of $P_{n-1}^\ast$ is \emph{irreducible} with respect to $P$ if there is no rewriting step of $P$ with source $u$.
A \emph{rewriting sequence of $P$ of length $k$} is an $(n-1)$-composition  
\[
f_1 \comp_{n-1} f_2 \comp_{n-1} \ldots \comp_{n-1} f_k
\]
in the free $n$-category~$P_n^\ast$, where the $f_i$ are rewriting steps of $P$.
If there exists such a rewriting sequence, we say that the $(n-1)$-cell $s_{n-1}(f_1)$ \emph{rewrites} to the $(n-1)$-cell $t_{n-1}(f_k)$. 
A \emph{zig-zag sequence of $P$ of length $k$} is an $(n-1)$-composition
\[
f_1^{\epsilon_1} \comp_{n-1} f_2^{\epsilon_2} \comp_{n-1} \ldots \comp_{n-1} f_k^{\epsilon_k}
\]
in the free $(n,n-1)$-category~$\tck{P}_n$, where the $f_i$ are rewriting steps of $P$, and $\epsilon_1,\ldots, \epsilon_k\in\{-1,1\}$, and which is
reduced with respect the rules $f\star_{n-1} f^- \fl 1$, for $n$-cells $f$ in $P_n^\ast$.

The rewriting steps of $P$ define an abstract rewriting system on the
set of parallel $(n-1)$-cells of the free $n$-category $P_n^\ast$, whose binary relation, denoted by $\ars{P_n}$, is defined by $u\ars{P_n} u'$ if there exists a rewriting step of $P$ that
reduces $u$ to $u'$.

\subsubsection{Remark}
\label{Rem:Contexts}
Given a cellular extension $\Gamma$ of an $n$-category $\Cr$, we also
denote by $\rrs{\Gamma}$ the set of \emph{cells of~$\Gamma$ in
  context}, that is the set of $(n+1)$-cells of the form

\[
f_{n}\comp_{n-1}\ldots \comp_2 (f_2\comp_1(f_1\comp_0 \alpha \comp_0 g_1)\comp_1 g_2) \comp_2 \ldots \comp_{n-1} g_{n},
\]
where $f_i, g_i$ are $i$-cells of $\Cr$ for $0\leq i \leq n$, and
$\alpha \in \Gamma$.  Recall from~{\cite[Prop. 2.1.5]{GuiraudMalbos09}}, that any $(n+1)$-cell~$\gamma$ in the
free $(n+1)$-category $\Cr[\Gamma]$ can be written as an
$n$-composition 
\[
\gamma = \gamma_1 \comp_n \gamma_2 \comp_{n} \ldots \comp_{n} \gamma_k,
\]
where the $\gamma_i$ are $(n+1)$-cells of $\rrs{\Gamma}$, using the algebraic laws
of higher categories, most notably the interchange laws.

\subsubsection{Rewriting properties of an $n$-polygraph}
\label{SSS:RewritingPropertiesPolygraphs}
The rewriting properties of an $n$-polygraph $P$ are those of the reduction relation $\ars{P_n}$. In particular, an $n$-polygraph $P$ is \emph{terminating} if there
is no infinite rewriting sequences with respect to $\ars{P_n}$.

A \emph{branching} of an $n$-polygraph $P$ is an unordered pair $(f,g)$ of rewriting sequences of $P$ such that $s_{n-1}(f) = s_{n-1}(g)$.
Such a branching is \emph{local} when $f$ and $g$ are rewriting steps.
We say that $P$ is \emph{confluent} (resp. \emph{locally confluent}) if for any branching (resp. local branching) $(f,g)$ there exist rewriting sequences $f'$ and $g'$ of $P$ with $t_{n-1}(f') = t_{n-1}(g')$ such that the compositions $f \star_{n-1} f'$ and $g \star_{n-1} g'$ are defined, as illustrated in the diagram
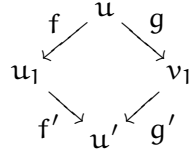
\[
\xymatrix@R=1.1em @C=1.1em{
&
u
\ar[dl] _-{f}
\ar[dr] ^-{g}
&
\\
u_1
\ar[dr] _-{f'}
&
&
v_1
\ar[dl] ^-{g'}
\\
&
u'
& 
}
\]
The \emph{source} of a branching $(f,g)$ is the common $(n-1)$-source $u$ of $f$ and $g$.
We say that $P$ is \emph{Church-Rosser} if for any zig-zag sequence $h$ of $P$ there exist rewriting sequences $h'$ and $k'$ of $P$ as in the  diagram
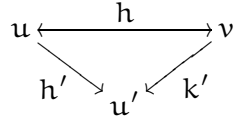
\[
\xymatrix@R=1.5em{
u
\ar@{<->}[rr] ^-{h}
\ar[dr] _-{h'}
&&
v
\ar[dl] ^-{k'}
\\
&
u'
&
}
\]

\subsubsection{Example: one-dimensional polygraphs} 
One-dimensional polygraphs are models of abstract rewriting systems.
Recall that an \emph{abstract rewriting system}
$\Ar=(X,\{ \fl_i \}_{i\in I})$ consists of a set $X$ and a family
${\fl}=\{ \fl_i \}_{i\in I}$ of binary relations on $X$, that is,
$\mathop{\fl_i }\subseteq X \times X$ for all
$i \in I$~\cite{Terese03}.  Then $\Ar$ can be described by a
$1$-polygraph $P=(P_0,P_1)$, whose set of generating $0$-cells is $X$, and whose set
of generating $1$-cells consists of
\[
u_{(x,y,i)} : x \fl y
\]
for all $x,y\in X$ and $i\in I$ such that $(x,y) \in \fl_i$.  If $I$
is a singleton, then $\Ar$ is a set $X$ together with a binary
relation $\fl$, and the underlying directed graph of the free
$1$-category $P_1^\ast$ is isomorphic to the reflexive and transitive
closure $\rstar$ of the relation $\fl$.  The underlying directed graph
of the $(1,0)$-category $\tck{P}_1$ is isomorphic to the symmetric
closure of the relation $\rstar$.

\subsubsection{Example: two-dimensional polygraphs} 
Two-dimensional polygraphs are models of string rewriting systems.  A
\emph{string rewriting system} is an abstract rewriting system on a
free monoid~\cite{BookOtto93}. It can be defined as a $2$-polygraph
$P=(P_0,P_1,P_2)$, where $P_0$ is a singleton, $P_1$ is an alphabet,
and the maps $s_0,t_0:P_1 \fl P_0$ are trivial. The free $1$-category
$P_1^\ast$ has one single $0$-cell. It thus isomorphic to the free
monoid generated by $P_1$, whose elements are the strings on
$P_1$. The cellular extension $P_2$ defines a binary relation on
strings on $P_1$, whose elements are the pairs
$(s_1(\alpha),t_1(\alpha))$, for $\alpha$ in $P_2$, and which are
rules of the string rewriting system. The binary relation $\fl_{P_2}$
is the rewrite relation generated by the set of rules.

\subsubsection{Example: three-dimensional polygraphs} 
Three-dimensional polygraphs are models of rewriting systems on free
$2$-categories. A \emph{two-dimensional diagrammatic rewriting system}
is a $3$-polygraph $P=(P_0,P_1,P_2,P_3)$, where the underlying
polygraph $(P_0,P_1)$ is the \emph{signature}, made of \emph{sorts} in
$P_0$ and \emph{generators} in $P_1$, the cellular extension $P_2$ is
the set of \emph{operators} with a finite number of inputs and
outputs, and the \emph{rules} relate \emph{two-dimensional diagrams}
made of $0$-compositions and $1$-compositions of operators in
$P_2$. Applications include term rewriting systems
for the explicit manipulation of variables in
terms~\cite{Guiraud06jpaa}, and  rewriting systems on monoidal
categories~\cite{GuiraudMalbos12mscs},
$2$-categories~\cite{Mimram14,GuiraudMalbos09} and linear
$2$-categories~\cite{Dupont2021}.

\subsection{Coherent confluence}
\label{SS:CoherenceConfluence}

We now define two notions of coherence of an
$n$-polygraph~$P$ with respect to a cellular extension $\Gamma$:
\begin{enumerate}
\item a \emph{vertical} one in which coherence cells, $(n+1)$-cells
  generated by $\Gamma$, have branchings as $n$-sources and 
  confluences as $n$-targets,
\item a \emph{horizontal} one in which coherence cells have
  rewriting sequences as $n$-sources and $n$-targets.
\end{enumerate}
A vertical approach has been used previously in  Kleene algebra, the
horizontal approach is the classical polygraphic approach.

\subsubsection{Coherent confluence}\label{SSS:CoherenceConfluence}
Let $P$ be an $n$-polygraph and $(f,g)$ be a branching of $P$.
A cellular extension $\Gamma$ of $P_n^\top$ is a \emph{confluence filler} for $(f,g)$ if there exist $n$-cells $f'$ and $g'$ in $P_n^\ast$, and two $(n+1)$-cells~$\alpha$ and $\alpha'$ in the $(n+1)$-category $\tck{P}_n[\Gamma]$ of the form $\alpha : f^-\star_{n-1} g \dfl f'\star_{n-1} (g')^-$ and $\alpha' :  g^- \star_{n-1} f \dfl g' \star_{n-1} (f')^-$:
\begin{equation}
\label{E:GammaConfluence}
\raisebox{1cm}{
\xymatrix@R=1.1em @C=1.1em{
&
u
\ar[dl] _-{f}
\ar[dr] ^-{g}
&
\\
u_1
\ar[dr] _-{f'}
&
&
v_1
\ar[dl] ^-{g'}
\\
&u'&
}}
\qquad
\raisebox{1cm}{
\xymatrix@R=1.1em @C=1.1em{
&
u
\ar@{<-}[dl] _-{f^-}
\ar[dr] ^-{g}
&
\\
u_1
\ar[dr] _-{f'}
&
&
v_1
\ar@{<-}[dl] ^-{(g')^-}
\\
&u'&
\ar@2 "1,2"!<0pt,-20pt>;"3,2"!<0pt,+20pt> ^-{\alpha} 
}}
\qquad
\raisebox{1cm}{
\xymatrix@R=1.1em @C=1.1em{
&
u
\ar[dl] _-{f}
&
\\
u_1
&
&
v_1
\ar[dl] ^-{g'}
\ar[ul] _-{g^-}
\\
&
u'
\ar[ul] ^-{(f')^-}
&
\ar@2 "1,2"!<0pt,-20pt>;"3,2"!<0pt,+20pt> ^-{\alpha'} 
}}
\end{equation}
The cellular extension $\Gamma$ is a \emph{confluence filler}
(resp. \emph{local confluence filler}) for the polygraph $P$ if
$\Gamma$ is a confluence filler for each of its branchings
(resp. local branchings).

Let $h$ be an $n$-cell in $P_n^\top$. The cellular extension $\Gamma$ is a \emph{Church-Rosser filler} for $h$ if there exist $n$-cells~$h'$ and $k'$ in $P_n^\ast$ and an $(n+1)$-cell $\alpha$ in the $(n+1)$-category $\tck{P}_n[\Gamma]$ of the form $\alpha : h \dfl h'\star_{n-1} {k'}^-$:
\begin{equation}
\label{E:GammaConfluence21}
\raisebox{0.5cm}{
\xymatrix@R=1.5em{
u
\ar[dr] _-{h'}
\ar@{<->}[rr] ^-{h}
&
&
v
\ar[dl] ^-{k'}
\\
&u'&
}
\qquad\qquad
\xymatrix@R=1.5em{
u
\ar[dr] _-{h'}
\ar@{<->}[rr] ^-{h}
&
&
v
\\
&
u'
\ar[ur] _-{(k')^-}
&
\ar@2 "1,2"!<+0pt,0pt>;"2,2"!<0pt,0pt> ^-{\alpha} 
}
}
\end{equation}
The cellular extension $\Gamma$ is a \emph{Church-Rosser filler} for
an $n$-polygraph $P$ if it is a Church-Rosser filler of every $n$-cell
in $P_n^\top$.

\subsubsection*{Remarks}
The $(n+1)$-cells $\alpha$ and $\alpha'$ in the definitions above are
$n$-compositions of $(n+1)$-cells of $\rrs{\Gamma}$ as defined in
Remark~\ref{Rem:Contexts}. Whiskering the $(n+1)$-cells $\alpha$ and
$\alpha'$ in~\eqref{E:GammaConfluence} yields $(n+1)$-cells
\begin{align*}
\beta := (g^-\star_{n-1} f) \star_{n-1} \alpha \star_{n-1} (g'\star_{n-1} (f')^-) : g' \star_{n-1}(f')^- 
&\fl g^- \star_{n-1} f,\\
\beta' := (f^-\star_{n-1} g) \star_{n-1} \alpha' \star_{n-1} (f'\star_{n-1} (g')^-) : f' \star_{n-1}(g')^- &\fl f^- \star_{n-1} g,
\end{align*}
as in the diagrams
\begin{equation}
\label{E:GammaConfluence22}
\raisebox{1cm}{
\xymatrix@R=1.1em @C=1.1em{
&
u
\ar@{<-}[dl] _-{f^-}
\ar[dr] ^-{g}
&
\\
u_1
\ar[dr] _-{f'}
&
&
v_1
\ar@{<-}[dl] ^-{(g')^-}
\\
&u'&
\ar@{<-}@2 "1,2"!<0pt,-20pt>;"3,2"!<0pt,+20pt> ^-{\beta'} 
}}
\quad\quad
\raisebox{1cm}{
\xymatrix@R=1.1em @C=1.1em{
&
u
\ar[dl] _-{f}
&
\\
u_1
&
&
v_1
\ar[dl] ^-{g'}
\ar[ul] _-{g^-}
\\
&
u'
\ar[ul] ^-{(f')^-}
&
\ar@{<-}@2 "1,2"!<0pt,-20pt>;"3,2"!<0pt,+20pt> ^-{\beta} 
}}
\end{equation}

\begin{thm}[Church-Rosser coherent filler lemma]
\label{T:CoherentCRARSFiller}
Let $P$ be an $n$-polygraph. A cellular extension~$\Gamma$ of
$P_n^\top$ is a confluence filler for $P$ if, and only if, $\Gamma$ is
a Church-Rosser filler for~$P$.
\end{thm}
\begin{proof}
  First suppose $\Gamma$ is a Church-Rosser filler for $P$. Then, for
  any branching $(f,g)$, the composites $f^- \star_{n-1} g$ and
  $g^- \star_{n-1} f$ are $n$-cells of $P_n^\top$, and $\Gamma$ is
  thus a Church-Rosser filler for them. This yields the cells $\alpha$
  and $\alpha'$ as in~\eqref{E:GammaConfluence}, and so $\Gamma$ is a
  confluence filler for $P$.
  
  Conversely, suppose $\Gamma$ is a confluence filler for $P$ and let
  $f$ be an $n$-cell of $P_n^\top$. We prove by induction on the
  length of $f$ that $\Gamma$ is a Church-Rosser filler for $f$.  This
  shows that $\Gamma$ is a Church-Rosser filler for~$P$.  For $f$ of
  length $0$ or $1$, $f$ is clearly $\Gamma$-confluent, since it
  suffices to take an identity $(n+1)$-cell. So suppose every $n$-cell
  of length $i \geq 2$ is $\Gamma$-confluent and that $f$ is of length
  $i+1$. Then $f=f_1 \star_{n-1} f_2$ with $f_1 : u \fl u_1$ in
  $P_n^\top$ of length $i$ and $f_2$ of length $1$ in $P_n^\ast$ is
  either of the form $v \fl u_1$ or $u_1 \fl v$. By the induction
  hypothesis, there exist $n$-cells $h$ and $k$ in $P_n^\ast$, and an
  $(n+1)$-cell $\alpha$ in $\tck{P}_n[\Gamma]$ such that
  $\alpha : f \Rightarrow h \star_{n-1} k^-$. If $f_2 : u_1 \fl v$,
  there exist $n$-cells $k'$ and $f''$ in $P_n^\ast$, and an
  $(n+1)$-cell $\beta$ in $\tck{P}_n[\Gamma]$ as shown in
  diagram~\eqref{E:ChurchRosserDiagramFiller} since $\Gamma$ is a
  confluence filler for $P$. Thus
  $(\alpha \star_{n-1} f_2) \star_n (h \star_{n-1}\beta)$ is a
  Church-Rosser filler for $f$.

\begin{equation}
\label{E:ChurchRosserDiagramFiller}
\raisebox{1cm}{
\xymatrix@R=3em @C=5em{
u
\ar@{<->}[rr] ^-{f_1} _-{}="src1"
\ar[dr] _-{h}
&&
u_1
\ar[r] ^-{f_2}
&
v
\\
&
u'
\ar[ur] |-{k^-}
\ar[r] _-{k'}
&
u''
\ar[ur] _-{{f''}^-}
&
\ar@2 "1,2"!<0pt,-8pt>;"2,2"!<0pt,+15pt> ^-{\alpha}
\ar@2 "1,3"!<0pt,-12pt>;"2,3"!<0pt,+15pt> ^-{\beta}
}}
\end{equation}
Otherwise, if $f_2 : v \fl u_1$, the $(n+1)$-cell
$\alpha \star_{n-1} f_2^-$ is a Church-Rosser filler for $f$:
\begin{equation}
\label{E:ChurchRosserDiagramFillerbis}
\raisebox{1cm}{
\xymatrix@R=3em @C=5em{
u
\ar@{<->}[rr] ^-{f_1} _-{}="src1"
\ar[dr] _-{h}
&&
u_1
\ar[r] ^-{(f_2)^-}
&
v
\\
&
u'
\ar[ur] |-{k^-}
&
&
\ar@2 "1,2"!<0pt,-8pt>;"2,2"!<0pt,+15pt> ^-{\alpha}
}}
\end{equation}
\end{proof}

\begin{thm}[Coherent Newman filler lemma]
\label{T:NewmanCoherentFiller}
Let $P$ be a terminating $n$-polygraph and $\Gamma$ a cellular
extension of $P_n^\top$. Then $\Gamma$ is a local confluence filler for $P$
if, and only if, $\Gamma$ is a confluence filler for $P$.
\end{thm}

\begin{proof}
  First observe that if $\Gamma$ is a confluence filler for $P$, then
  it is also a local confluence filler for $P$ since local branchings
  are branchings.

  Now suppose $\Gamma$ is a local confluence filler for $P$. We prove
  by Noetherian induction that, for every $(n-1)$-cell $u$
  of~$P_n^\ast$, $\Gamma$ is a confluence filler for every branching
  of $P$ with source $u$.  For the base case, if~$u$ is irreducible
  for $P$, then~$(1_u,1_u)$ is the only branching with source~$u$, and
  it is $\Gamma$-confluent, taking the $(n+1)$-cell $1_{1_u}$ in $\tck{P}_n[\Gamma]$.

  For the induction step, suppose~$u$ is a reducible $(n-1)$-cell
  of~$P_n^\ast$ and $\Gamma$ a confluence filler for every branching
  with source an $(n-1)$-cell~$u'$ such that $u$ rewrites to $u'$.
  Let~$(f,g)$ be a branching of~$P$ with source~$u$. If one of~$f$
  or~$g$ is an identity, $f$ say, then $\Gamma$ is a confluence filler
  for~$(f,g)$ by considering the $(n+1)$-cells $1_g$ and
  $1_{g^-}$ in $\tck{P}_n[\Gamma]$. 
  Otherwise, if the $n$-cells $f$ and $g$ are not
  identities, then we may write $f=f_1\comp_{n-1} f_2$ and
  $g=g_1\comp_{n-1} g_2$, where $g_1, f_1$ are rewriting steps and
  $g_2,f_2$ are $n$-cells of $P_n^\ast$.  Since $\Gamma$ is a local
  confluence filler for $P$, there exist $n$-cells $f_1',g_1'$ in
  $P_n^\ast$, and an $(n+1)$-cell $\alpha$ in $P_n^\ast[\Gamma]$ as in
  the diagram~\eqref{E:DiamondProofNewmanFiller}.  We can apply the
  induction hypothesis to the branching $(f_2,f_1')$, which yields
  $n$-cells $f_2',h$ in $P_n^\ast$ and an $(n+1)$-cell $\beta$ in
  $P_n^\ast[\Gamma]$ as in the
  diagram~\eqref{E:DiamondProofNewmanFiller}. Finally, we can apply
  the induction hypothesis again to the branching
  $(g_1'\star_{n-1} h, g_2)$, which yields $n$-cells $k$ and $g_2'$ in $P_n^\ast$
  and an $(n+1)$-cell $\gamma$ in $P_n^\ast[\Gamma]$ as
  in~\eqref{E:DiamondProofNewmanFiller}.

\begin{equation}
\label{E:DiamondProofNewmanFiller}
\raisebox{2.7cm}
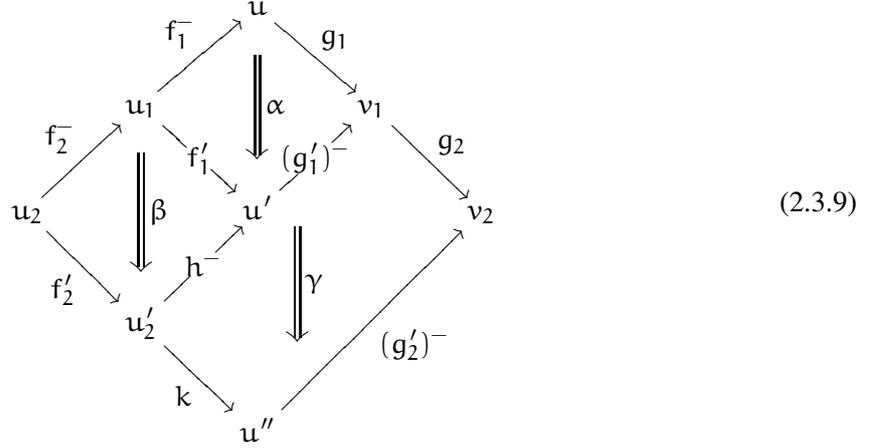
{
\xymatrix@R=2.3em @C=2.3em{
&&
u
\ar@{<-}[dl] _-{f_1^-}
\ar[dr] ^-{g_1}
&&
\\
&\ar[dr] |-{f_1'}
\ar@{<-}[dl] _-{f_2^-}
u_1&&
\ar@{<-}[dl] |-{(g_1')^-}
\ar[dr] ^-{g_2}
v_1
&
\\
u_2
\ar[dr] _-{f_2'}
&&
u'
\ar@{<-}[dl] |-{h^-}
&&
v_2
\ar@{<-}[ddll] ^-{(g_2')^-}
\\
&
u_2'
\ar[dr] _-{k}
&&&
\\
&&u''&&
\ar@2 "1,3"!<0pt,-18pt>;"3,3"!<0pt,+20pt> ^-{\alpha}
\ar@2 "2,2"!<0pt,-18pt>;"4,2"!<0pt,+20pt> ^-{\beta}
\ar@2 "3,3"!<+15pt,0pt>;"5,3"!<+15pt,+35pt> ^-{\gamma}
}}
\end{equation}
The $n$-composition
\begin{equation}
\label{E:DiamondProofNewmanEqn1}
\delta = ((
(f_2^- \comp_{n-1} \alpha )
\comp_n
(\beta \comp_{n-1} (g_1')^-)
)
\comp_{n-1} g_2)
\comp_n
(f_2'\comp_{n-1} \gamma)
\end{equation}
is an $(n+1)$-cell in $P_n^\ast[\Gamma]$ with source $f^-  \comp_{n-1}
g$ and target $f_2' \comp_{n-1} k \comp_{n-1} (g_2')^-$. We can
similarly find an $(n+1)$-cell $\delta'$ with source $g^- \comp_{n-1}
f$ and with target a confluence. 
As a consequence, $\Gamma$ is a confluence filler
for $P$, which proves the result.
\end{proof}

\subsubsection{Remark}\label{Rem:HDRadvantages}
Readers familiar with abstract rewriting may notice that the proofs of
Theorems~\ref{T:CoherentCRARSFiller} and~\ref{T:NewmanCoherentFiller}
are similar to the classical ones for abstract rewriting
systems. Indeed, forgetting the $(n+1)$-dimensional coherence cells
and look only at their $n$-dimensional borders in
\eqref{E:ChurchRosserDiagramFiller},
\eqref{E:ChurchRosserDiagramFillerbis} and
\eqref{E:DiamondProofNewmanFiller} yields precisely the diagrams used
to prove the $1$-dimensional results for abstract rewriting
systems. The higher-dimensional approach is thus consistent with the
abstract case while offering several advantages. First, using explicit
witnesses for confluence allows for a constructive formulation of
classical results using normalisation strategies.  Furthermore, as the
higher-dimensional cells may be considered as rewriting systems in
their own right, and as the procedures described above work in any
dimension, higher rewriting provides a constructive method for
calculating resolutions and cofibrant replacements of algebraic
structures.  Another advantage is that we work directly on rewrite
sequences instead of relations.

\subsection{\texorpdfstring{$\Gamma$}{Gamma}-confluence and filling}
\label{SS:GammaConfluenceAndFilling}

Recall from~\cite{GuiraudHoffbeckMalbos19} that, for any $n$-polygraph
$P$ and a cellular extension of $P_n^*$, we say that $P$ is
\emph{$\Gamma$-confluent} (resp. \emph{$\Gamma$-locally confluent}) if
for every branching (resp. local branching) $(f,g)$ of $P$ there exist
$n$-cells $f',g'$ in the free $n$-category $P_n^\ast$, and an
$(n+1)$-cell $\alpha : f \star_{n-1} f' \dfl g \star_{n-1} g'$ in the
free $(n+1,n)$-category $P_n^\ast(\Gamma)$ as in the diagram
\begin{equation}
\label{E:GammaConfluence3}
\xymatrix@R=1.1em @C=1.1em{
&
u
\ar[dl] _-{f}
\ar[dr] ^-{g}
&
\\
u_1
\ar[dr] _-{f'}
&
&
v_1
\ar[dl] ^-{g'}
\\
&u'&
\ar@2 "2,1"!<+20pt,0pt>;"2,3"!<-20pt,0pt> ^-{\alpha} 
}
\end{equation}
We say that $P$ is \emph{$\Gamma$-Church-Rosser} if for every
$n$-cell $h$ of $P_n^\top$ there exist $n$-cells $h'$ and $k'$ in the free
$n$-category $P_n^*$ and an $(n+1)$-cell $\alpha: h \star_{n-1} h' \dfl k'$ in the free $(n+1,n)$-category $\tck{P}_n(\Gamma)$ as in the  diagram
\begin{equation}
\label{E:GammaConfluence4}
\xymatrix@R=1.5em{
u
\ar[dr] _-{h'}
\ar@{<->}[rr] ^-{h}
&
&
v
\\
&
u'
\ar[ur] _-{(k')^-}
&
\ar@2 "2,1"!<+25pt,+17pt>;"2,3"!<-25pt,+17pt> ^-{\alpha} 
}
\end{equation}

Theorems~\ref{T:CoherentCRARSFiller} and~\ref{T:NewmanCoherentFiller}
were formulated in terms of fillers above. Now we express them using
$\Gamma$-confluence.

\begin{thm}[Church-Rosser coherent lemma]
\label{T:CoherentCRARS}
Let $P$ be an $n$-polygraph and $\Gamma$ a cellular extension of
$P_n^\ast$. 
The polygraph $P$ is $\Gamma$-confluent if, and only if, it is
$\Gamma$-Church-Rosser.
\end{thm}
\begin{proof}

  The proof is similar to that of Theorem~\ref{T:CoherentCRARSFiller},
  but with $(n+1)$-cells oriented horizontally in the induction step,
  as pictured in the following diagram:
\begin{equation}
\label{E:ChurchRosserDiagram}
\raisebox{1cm}{
\xymatrix@R=3em @C=3em{
u
\ar@{<->}[rr] ^-{f}
\ar[dr] _-{h}
&&
v
\ar[r] ^-{f'}
\ar[dl] |-{k}
&
\ar[dl] ^-{f''}
\\
&
u'
\ar[r] _-{k'}
&
v'
&
\ar@2 "1,1"!<+35pt,-20pt>;"1,3"!<-35pt,-20pt> ^-{\alpha}
\ar@2 "1,2"!<+40pt,-20pt>;"1,4"!<-30pt,-20pt> ^-{\beta}
}}
\end{equation}
The composite $(\alpha \star_{n-1} k') \star_n (f\star_{n-1} \beta)$ makes the $n$-cell $f$ $\Gamma$-confluent. 
\end{proof}

\begin{thm}[Coherent Newman lemma]
\label{T:NewmanCoherent}
Let $P$ be a terminating $n$-polygraph and $\Gamma$ a cellular extension of $P_n^\ast$. The polygraph $P$ is locally $\Gamma$-confluent if, and only if, it is $\Gamma$-confluent.
\end{thm}
\begin{proof}

The proof is similar to that of Theorem~\ref{T:NewmanCoherentFiller}, but with the following induction diagram:
\begin{equation}
\label{E:DiamondProofNewman}
\raisebox{2.7cm}{
\xymatrix@R=1.7em @C=1.7em{
&&
u
\ar[dl] _-{f_1}
\ar[dr] ^-{g_1}
&&
\\
&\ar[dr] |-{f_1'}
\ar[dl] _-{f_2}
u_1&&
\ar[dl] |-{g_1'}
\ar[dr] ^-{g_2}
v_1
&
\\
u_2
\ar[dr] _-{f_2'}
&&
u'
\ar[dl] |-{h}
&&
v_2
\ar[ddll] ^-{g_2'}
\\
&
u_2'
\ar[dr] _-{k}
&&&
\\
&&u''&&
\ar@2 "2,2"!<+25pt,0pt>;"2,4"!<-25pt,0pt> ^-{\alpha}
\ar@2 "3,1"!<+25pt,0pt>;"3,3"!<-25pt,0pt> ^-{\beta}
\ar@2 "4,2"!<+45pt,+20pt>;"4,4"!<-5pt,+20pt> ^-{\gamma}
}}
\end{equation}
The  $n$-composition 
\begin{equation}
\label{E:DiamondProofNewmanEqn2}
\delta = ((
(f_1\comp_{n-1} \beta )
\comp_n
(\alpha\comp_{n-1} h)
)
\comp_{n-1} k)
\comp_n
(g_1\comp_{n-1} \gamma)
\end{equation}
is then an $(n+1)$-cell in $P_n^\ast(\Gamma)$ with source
$f \comp_{n-1} (f_2' \star_{n-1} k)$ and target $g \comp_{n-1}g_2'$,
proving the result.
\end{proof}

For $\Gamma=\Sph(P_n^\ast)$, (local) $\Gamma$-confluence
(resp. $\Gamma$-Church-Rosser) coincides with of (local) confluence
(resp. Church-Rosser) of $P$ as defined
in~\eqref{SSS:RewritingPropertiesPolygraphs}.
Theorems~\ref{T:NewmanCoherent} and \ref{T:CoherentCRARS} correspond
to Newman's lemma and the Church-Rosser theorem~\cite{Newman42}, see
also~\cite{Huet80}.

\subsubsection{Remarks}
In this section, we have defined a vertical and a horizontal notion of
coherence of an $n$-polygraph~$P$ with respect to a cellular extension
$\Gamma$. The vertical notion requires inverses of $n$-cells, that is,
$\Gamma$ is a cellular extension of $P_n^\top$. The proofs of
Theorems~\ref{T:CoherentCRARSFiller} and~\ref{T:NewmanCoherentFiller}
do not need inverses of $(n+1)$-cells. The horizontal notion, by
contrast, does not need inverses of $n$-cells, that is, we consider
cellular extensions of $P_n^*$, but only inverses of $(n+1)$-cells are
needed to prove Theorems~\ref{T:CoherentCRARS}
and~\ref{T:NewmanCoherent}.  In the vertical approach, the proofs thus
take place in $\tck{P}_n[\Gamma]$ whereas, in the horizontal one, they
take place in $P_n^\ast(\Gamma)$. Furthermore, in the first approach,
we specify two filler cells $\alpha$ and $\alpha'$ as depicted in
diagram~\eqref{E:GammaConfluence} for each branching
$(f,g)$. Branchings are unordered pairs, we must therefore account for
both cases. This is another reason why we require inverses of
$(n+1)$-cells in the horizontal approach.

  In the remainder of this article, we exclusively consider the
  vertical approach to paving diagrams with higher-dimensional cells.

\section{Higher modal Kleene algebras}
\label{S:HigherDimKleene}

In this section we introduce higher globular modal Kleene algebras. In
its first subsection, we list the axioms of \emph{modal Kleene
  algebra}~\cite{DesharnaisStruth11} and two of its main models.  Its
relational model provides the original intuition for defining modal
operators based on relational domain and codomain operations over
Kripke frames.  Its path model, which can be defined over any graph,
forms the basis for using modal Kleene algebras in higher
rewriting. We then define $n$-dimensional dioids and equip these with
domain, codomain and star operations to obtain \emph{modal $n$-Kleene
  algebras}. Finally, we construct a higher path algebra associated to
an $n$-polygraph with a cellular extension $\Gamma$ as a model of this
structure.

\subsection{Modal Kleene algebras}
\label{SS:ModalKleeneOne}

\subsubsection{Semirings}

A \emph{semiring} is a structure $(S, + , 0, \cdot, 1)$ made of a set
$S$ and two binary operations~$+$ and $\cdot$ such that $(S,+,0)$ is a
commutative monoid, $(S,\cdot,1)$ is a monoid whose
\emph{multiplication operation}~$\cdot$ distributes over the
\emph{addition operation} $+$, from the left and right, and $0$ is a
left and right zero of multiplication.  A \emph{dioid} is a semiring
$S$ in which addition is idempotent: $x+x=x$ for all $x\in S$. In this
case, $(S,+,0)$ is a semilattice with partial order defined by
\begin{equation}
\label{E:OrderDiod}
x\leq y \quad \Leftrightarrow \quad x + y = y,
\end{equation}
for all $x,y\in S$, with respect to which addition and multiplication
are order-preserving and $0$ is minimal.  We will often denote
multiplication simply by juxtaposition.

A \emph{bounded distributive lattice} is a dioid $(S,+,0,\cdot,1)$,
whose multiplication $\cdot$ is commutative and idempotent, and
$x \leq 1$, for every $x\in S$.

\subsubsection{Domain semirings}
\label{SSS:DomainSemiringsOne}
A \emph{domain semiring}~\cite{DesharnaisStruth11} is
a dioid $(S, + , 0,\cdot, 1)$ equipped with a \emph{domain
  operation} $\dom : S \rightarrow S$ that satisfies the following
five axioms. For all $x,y\in S$,
\begin{enumerate}[{\bf i)}]
\item $x \leq \dom(x)x$,
\item $\dom(xy) = \dom(xd(y))$,
\item $\dom(x)\leq 1$,
\item $\dom(0) = 0$,
\item $\dom(x+y) = \dom(x) + \dom(y)$.
\end{enumerate}

These structures are called domain \emph{semirings} and not domain
\emph{dioids} because semirings equipped with a domain operation are
automatically idempotent~\cite{DesharnaisStruth11}.

Intuitions for the domain axioms are given in
  Examples~\ref{RelationModelOne} and \ref{PathModelOne} below. In the
  first, we explain that the domain of a binary relation, which
  models the set of all elements that it relates to another element of
  the underlying set, satisfies the domain semiring axioms. The second
  example shows that the algebra of sets of paths over a digraph or
  quiver, represented by a $1$-polygraph, satisfies the domain
  semiring axioms. The domain of a set of paths then corresponds to
  the set of all sources of paths in the set.

Consequences of the domain semiring axioms include the fact that the
image of $S$ under $\dom$ is precisely the set of fixpoints of $\dom$,
that is, 
$$S_\dom : = \{x\in S\mid \dom(x) = x\}= \dom(S),$$ 
and that $S_\dom$ forms a distributive lattice with $+$ as join and
$\cdot$ as meet, bounded by $0$ and $1$. It contains the largest
Boolean subalgebra of $S$ bounded by $0$ and $1$.  We henceforth write
$p,q,r,\dots$ for elements of $S_\dom$ and refer to $S_\dom$ as the
\emph{domain algebra} of $S$.  In particular, $S_\dom$ is a
subsemiring of $S$ in the sense that its elements satisfy the semiring
axioms, $0$ and $1$ are in the set, and the set is closed with respect
to $\cdot$ and $+$.

In the relational model of domain semirings, the set $S_d$ consists of
the set of all relations included in the identity relation, called
\emph{subidentities}. In the path model, it consists of subsets of the
set of all paths of length $0$. In both cases, the distributive
sublattices form Boolean algebras.

Further properties of domain semirings include
\[
\dom(0)=0,
\qquad
\dom{(px)} = p\dom(x),
\qquad
x\le y\Rightarrow \dom(x)\le \dom(y), 
\]
for all $x,y \in S_d$, and $\dom$ commutes with all existing
sups~\cite{DesharnaisStruth11}.

\subsubsection{Boolean domain semirings}\label{SSS:BooleanSemiringsOne}
A limitation of domain semirings is that Boolean complementation in
$S_d$ cannot be expressed; these structures admit chains as
models~\cite{DesharnaisStruth11}. Yet complementation is desirable for
at least two reasons: It reflects the Boolean nature of the path
models, in which we are interested, more faithfully.  It also allows
us to define a modal box operator from the modal diamond, built using
domain, via standard De Morgan duality, see
(\ref{SSS:ModalitiesKleeneOne}). We need both Boolean domain algebras
and the box-diamond duality in the proof of coherent Newman's lemma in
Section~\ref{SS:CoherentNewman}.

To enforce Boolean domain algebras, it is standard to axiomatise a
notion of antidomain that abstractly describes those elements that are
\emph{not} in the domain of a particular element. The antidomain
  of a relation, for instance, models the set of all elements that are
  not related to any other element of the underlying set; the
  antidomain of a set of paths corresponds to the set of all vertices
  of the underlying graph that are not a source of any path in the
  set.

  A \emph{Boolean domain semiring}~\cite{DesharnaisStruth11} is a
  dioid $(S, + , 0,\cdot , 1)$ equipped with an \emph{antidomain
    operation} $\adom : S \rightarrow S$ that satisfies, for all
  $x,y\in S$:

\begin{enumerate}[{\bf i)}]
\item $\adom(x)x = 0$, 
\item $\adom(xy) \leq \adom(x \, \adom^2 (y) )$, 
\item $\adom^2(x) + \adom(x) = 1$.
\end{enumerate}
As the antidomain operation is, implicitly, the Boolean complement of
the domain operation, we have $\dom = \adom^2$. Hence we recover a
domain semiring: $\dom$ satisfies the domain semiring axioms. In the
presence of $\adom$, the subalgebra $S_\dom$ of all fixpoints of
$\dom$ in $S$ is now the greatest Boolean algebra in $S$ bounded by
$0$ and $1$, and $S_\dom = \adom(S)$ and $\adom$ acts indeed as
Boolean complementation on $S_\dom$. We therefore write $\neg$ for the
restriction of $\adom$ to $S_\dom$.

\subsubsection{Modal semirings}\label{SSS:modal-semirings}
We denote the \emph{opposite} of a semiring $S$, in which the order of
multiplication has been reversed, by $\op{S}$.  It is once again a semiring. 
A \emph{codomain} (resp. \emph{Boolean codomain}) \emph{semiring} is a
semiring equipped with a map $\cod:S \fl S$ (resp. $\acod:S \fl S$) such that
$(\op{S} , \cod)$ (resp. $(\op{S} , \acod)$) is a domain
(resp. Boolean domain) semiring.

As expected, the codomain operation models the domain of the
  converse relation in the relational model, and in the path model the
  set of all targets of paths in a given set of paths.

Consider a semiring equipped with a domain and a codomain operation.
The domain and codomain axioms alone do not imply that
$S_\dom = S_\cod$, let alone the compatibility properties
\begin{equation}
\label{E:coherenceModalSemiring}
\dom(\cod(x)) = \cod(x),
\qquad
\cod(\dom(x)) = \dom(x),
\end{equation}
for every $x \in S$.  Indeed, consider the domain and range semiring
$S=(\{a\},+,0,\cdot,1,\dom,\cod)$ with addition defined by $0<a<1$,
multiplication by $a^2=a$, domain by $\dom(a) = 1$ and codomain by
$\cod(a) = a$. Then $S_\dom=\{0,1\}\neq \{0,a,1\}=S_\cod$ and
$\dom(\cod(a)) = 1\neq a = \cod(a)$, but $\cod\circ \dom = \dom$
fails in the opposite semiring.

A \emph{modal semiring} $S$~\cite{DesharnaisStruth11} is a domain
semiring and a codomain semiring that satisfies the compatibility
properties~\eqref{E:coherenceModalSemiring}.  Boolean domain semirings
that are also Boolean codomain semirings are called \emph{Boolean
  modal semirings}.  In this case, maximality of $S_\dom$ and
$S_\cod=\{x\in S\mid \cod(x)=x\}$ forces the domain and range algebra
of $S$ to coincide, so that the extra axioms
\eqref{E:coherenceModalSemiring} are unnecessary. We provide a formal
proof, as this fact has so far been overlooked in the literature.
\begin{lem}
  In every Boolean modal semiring the compatibility
  properties~\eqref{E:coherenceModalSemiring} hold.
\end{lem}
\begin{proof}
Suppose $S$ is a Boolean modal semiring and let $x$ in $S$. Then
 \begin{align*}
    \dom(\cod(x)) 
&= (\acod(x) + \cod(x)) \dom(\cod(x)) \\
&= \acod(x) \dom(\cod(x)) 
+ \cod(x) \dom(\cod(x)) (\acod(x) +\cod(x))\\
&= 0+ \cod(x) \dom(\cod(x))  \acod(x)
+\cod(x) \dom(\cod(x)) \cod(x)\\
&= 0+ \cod(x) \cod(x) = \cod(x)
  \end{align*}
  proves the first identity in~\eqref{E:coherenceModalSemiring}.

  In
  the third step, we have $\acod(x)\dom (\cod(x)) = 0$ because
  $\acod(x)\cod(x) = 0$ and $yz=0\Leftrightarrow yd(z)=0$ hold in any
  Boolean modal semiring.  In the fourth step,
  $\cod(x) \dom(\cod(x)) \acod(x) = 0$ because $\dom(\cod(x)) \le 1$
  and again $\acod(x)\cod(x) = 0$. Moreover
  $\cod(x) \dom(\cod(x)) \cod(x) = \cod(x) \cod(x)$ because
  $\dom(y)y=y$ holds in any modal semiring.

  The proof of the second
  identity in~\eqref{E:coherenceModalSemiring} follows by opposition.
\end{proof}

In Boolean modal semirings, $\dom(x)=x$ therefore implies
$\cod(x)=\cod(\dom(x)) =\dom(x)=x$, while $\cod(x)=x$ implies
$\dom(x)=x$ by opposition.  This forces that $S_\dom=S_\cod$, as desired.

\subsubsection{Modal Kleene algebras}
\label{SSS:mka}

A \emph{Kleene algebra} is a dioid $K$ equipped with a \emph{Kleene
  star} $(-)^* : K \fl K$ that satisfies, for all $x,y,z\in K$,

\begin{enumerate}
\item (\emph{unfold axioms}) $1 + xx^*  \leq  x^*$ and $1 + x^*x  \leq x^*$,
\item (\emph{induction axioms}) $z +  xy \leq y \Rightarrow  x^*z \leq y$ and 
$z + yx \leq y \Rightarrow  zx^* \leq y$.
\end{enumerate}

The axioms on the left are the opposites of those on the
right. Intuitively, the axioms for the Kleene star model a finite
iteration of an element $x$ as a least fixpoint. The first unfold
axiom, for instance, states that iterating $x$ either amounts to doing
nothing, that is, doing $1$, or doing $x$ once and then continuing the
iteration. As possibly infinite iterations would satisfy such unfold
laws, too, the induction laws filter out the least fixpoints of the
corresponding pre-fixpoint equations. More detailed explanations of
the induction laws can be found in the literature. In the relational
model, $(-)^\ast$ is the reflexive-transitive closure of a relation,
in the path model it captures the repetitive composition of paths in a
given set.

Useful consequences of Axioms {\bf i)} and {\bf ii)} include, for all
$x,y\in K$, and $i\in\mathbb{N}$,
\[
x^i\le x^\ast
\quad
x^\ast x^\ast = x^\ast
\quad
x^{\ast\ast} = x^\ast
\quad
x(yx)^\ast = (xy)x^\ast
\quad
(x+y)^\ast = x^\ast (yx^\ast)^\ast = (x^\ast y^\ast)^\ast,
\] 
where $x^i$ denotes the $i$-fold multiplication of $x$ with itself, as well as the quasi-identities 
\[
x\le 1\Rightarrow x^\ast = 1
\quad
x\le y\Rightarrow x^\ast \le y^\ast
\quad
xz\le zy\Rightarrow x^\ast z\le zy^\ast
\quad
zx\le yz\Rightarrow zx^\ast \le y^\ast z. 
\]
The \emph{Kleene plus}  $(-)^+ : K \fl K$ is defined as
$x^+ = xx^*$. It corresponds to the transitive closure operation
  in the relational model.

  The above notions of domain and codomain extend to Kleene algebras
  without any additional axioms. A \emph{(Boolean) modal Kleene
    algebra} is thus a Kleene algebra that is also a (Boolean)
  modal-semiring.

\subsubsection{Modal Operators}
\label{SSS:ModalitiesKleeneOne}

In our algebraic approach to higher rewriting, modalities allow
relating sets of higher-dimensional cells to their sets of
lower-dimensional source and target cells,
see~\eqref{SSS:GlobularSemiring}, and thus expressing the
forall/exists properties defining fillers and pasting conditions in
proofs of higher rewriting.

In the relational model of the $1$-dimensional case, $\fDia{x}p$
indicates the subset of the underlying set from which one may reach
the set $p$ along relation $x$, and $\bDia{x}p$ the set that one may
reach from $p$ along $x$. Similarly, $\fBox{x}p$ indicates the set
from which we must reach the set $p$ along $x$, and $\bBox{x}p$ the
set that we must reach from $p$ along $x$. Similar intuitions underlie
the path model of modal Kleene algebra, and these generalise to the
notions of higher paths and their relations expressed in the filler
properties and pasting conditions of higher rewriting. These
explanations motivate the following algebraic definitions.
 
Let $(S,+,0,\cdot,1,\dom,\cod)$ be a modal semiring.  For $x\in S$
and $p\in S_d$, we define the forward and backward modal diamond operators
\begin{equation}
\label{ModalitiesDefOne}
  \fDia{x}p = \dom(xp) \qquad\text{ and }\qquad \bDia{x}p=\cod(px). 
\end{equation}
When $S$ is a Boolean modal semiring, we additionally define the
forward and backward modal box operators
\begin{equation}
\label{eq:boxes}
  \fBox{x}p=\neg\fDia{x}(\neg p) \qquad \text{ and }\qquad
  \bBox{x}p=\neg\bDia{x}(\neg p).
\end{equation}
Beyond the intuitions given, these are modal operators in the sense of
Jónsson and Tarski's Boolean algebras with
operators~\cite{JonssonTarski51} because the identities
\[
\fDia{x}(p+q)=\fDia{x}p+\fDia{x}q, 
\qquad 
\fDia{x}0=0,
\qquad
\bDia{x}(p+q)=\bDia{x}p+\bDia{x}q, 
\qquad 
\bDia{x}0=0,
\]
hold, and dually
\[
\fBox{x}(pq)=\fBox{x}p+\fBox{x}q, 
\qquad 
\fBox{x}1=1,
\qquad
\bBox{x}(pq)=\bBox{x}p+\bBox{x}q, 
\qquad
\bBox{x}1=1.  
\]

It is easy to see that $\fDia{-}$ and $\bDia{-}$, as well as
$\fBox{-}$ and $\bBox{-}$ are related by opposition. In a (Boolean)
modal Kleene algebra, following J\'onsson and Tarski, this can be
expressed by the conjugation laws
\begin{equation*}
  \fDia{x}p\cdot q = 0 \Leftrightarrow p\cdot \bDia{x}q=0\qquad\text{
    and }\qquad\fBox{x}p+q=1\Leftrightarrow p+\bBox{x}q = 1.
\end{equation*}
In the relational model, it can be expressed explicitly using
  relational converse.

In a Boolean modal semiring, boxes and diamonds are related by De Morgan duality by their definition~\eqref{eq:boxes} and additionally by
\begin{equation}
\label{E:DeMorgan}
  \fDia{x}p=\neg\fBox{x}(\neg p)\qquad\text{ and }\qquad
  \bDia{x}p=\neg \bBox{x}(\neg p).
\end{equation}
Finally, boxes and diamonds are adjoints in Galois connections:
\[
  \fDia{x}p\le q \Leftrightarrow p\le \bBox{x}q\qquad \text{ and
  }\qquad \bDia{x}p\le q \Leftrightarrow p\le \fBox{x}q.
\]
As a consequence, diamonds preserve all existing sups in $S$,
whereas boxes reverse all existing infs to sups, and all modal operators
are order preserving. 
Finally, we mention the properties
$\fDia{xy}=\fDia{x}\circ \fDia{y}$,
$\bDia{xy}=\bDia{y}\circ \bDia{x}$,
$\fBox{xy}=\fBox{x}\circ \fBox{y}$ and 
$\bBox{xy}=\bBox{y}\circ \bBox{x}$. 

\subsubsection{Example: relation Kleene algebra}
\label{RelationModelOne}

Here we put our aforementioned intuitions on solid foundations. The
relational model of plain Kleene algebra has been the starting point
for Kleene-algebraic proofs of the Church-Rosser theorem of abstract
rewriting, that of modal Kleene algebra has motivated the
Kleene-algebraic proof of Newman's lemma.

For any set $X$, the structure
\[
(\mathcal{P}(X\times X) , \cup\, , \emptyset_X , \mathop{;}\, , 
\mathit{Id}_X, (-)^*)
\]
forms a Kleene algebra, the \emph{full relation Kleene algebra} over
$X$. The operation $\mathop{;}$ is  relational composition defined
by $(a,b)\in R\mathop{;} S$ if, and only if, $(a,c)\in R$ and
$(c,b)\in S$, for some $c\in X$.  The relation
$\mathit{Id}_X=\{(a,a)\mid a\in X\}$ is the identity relation on $X$
and $(-)^\ast$ is the reflexive transitive closure operation defined, for
$R^0=\mathit{Id}_X$ and $R^{i+1}=R\mathop{;} R^i$, by
\begin{equation*}
  R^\ast = \bigcup_{i\in\mathbb{N}} R^i.
\end{equation*}
The subidentity relations below $\mathit{Id}_X$ form the greatest
Boolean subalgebra between $\emptyset_X$ and $\mathit{Id}_X$. It is
isomorphic to the power set algebra $\mathcal{P}(X)$. Every subalgebra
of a full relation Kleene algebra is a \emph{relation Kleene algebra}.

The full relation Kleene algebra over $X$ extends to a
\emph{full relation
  Boolean modal Kleene algebra} over $X$ by defining, as expected, 
  \begin{equation*}
    \dom(R) = \{(a,a) \mid \exists b\in X.\ (a,b)\in
    R\}\qquad\text{ and }\qquad \cod(R) = \{(a,a)\mid  \exists b.\ (b,a)\in R\}.
  \end{equation*}
The domain algebra $\mathcal{P}(X\times X)_\dom$ equals the
    Boolean algebra of subidentity relations.

  The antidomain and anticodomain maps are then given by relative
  complementation $\adom(R) = \mathit{Id}_X\setminus\dom(R)$ and
  $\acod(R) = \mathit{Id}_X\setminus \cod(R)$ within the domain
  algebra. Finally, it is straightforward to check that the algebraic
  definitions of boxes and diamonds expand to their standard
  relational Kripke semantics:
  \begin{align*}
    \fDia{R}P &= \{(a,a)\mid \exists b\in X.\ (a,b)\in R\land (b,b) \in
    P\},\\
 \fBox{R}P&=\{(a,a)\mid \forall b\in X.\
    (a,b)\in R\Rightarrow (b,b)\in P\},
  \end{align*}
  and likewise for the backward modalities. This requires swapping
  $(a,b)$ to $(b,a)$ in the above expressions, which amounts to taking
  relational converse. 

  \subsubsection{Example: path Kleene algebras}\label{PathModelOne}

  The path model of modal Kleene algebra is a stepping stone towards
  polygraph models of higher Kleene algebras. Instead of a
  $1$-polygraph, we could speak of a directed graph or quiver.  So let
  $P^\ast$ be the free $1$-category generated by the $1$-polygraph
  $P =( P_0, P_1)$. Its elements are paths in $P$ to which we assign
  source and target maps $s_0$ and $t_0$ as well as a path 
  composition~$\star_0$ in the standard way. Then
  $(\mathcal{P}(P_1^*) , \cup , \emptyset, \odot, \uno , (-)^*)$ forms
  a Kleene algebra, the \emph{full path (Kleene) algebra} $K(P)$ over
  $P$.  Here, composition is defined as a complex product
\begin{equation*}
\phi\odot\psi = \{\;u\star_0v\,\mid\, u\in \phi\,\land\, v\in\psi\, \land\, t_0(u)=s_0(v)\;\}
\end{equation*}
for any $\phi,\psi\in\mathcal{P}(P_1^\ast)$, and $\uno$ is the set of
all identity arrows, or paths of length zero, of $P$. The Kleene star is defined as
\begin{equation*}
  \phi^\ast = \bigcup_{i\in\mathbb{N}} \phi^i
\end{equation*}
where $\phi^0=\uno$ and $\phi^{i+1} = \phi\odot\phi^i$. It
  models the repetitive composition of the paths in $\phi$ mentioned
  before. Every
subalgebra of the full path Kleene algebra over $P$ is a \emph{path
  Kleene algebra}. As in the case of relational Kleene algebras, the set of all
subidentities (subsets of $\uno$), the set of sets of identity arrows, forms a
Boolean subalgebra.

The full path algebra over $P$ extends to a \emph{full path Boolean
  modal Kleene algebra} over $P$ by defining
\begin{equation*}
  \dom(\phi) =  \{1_{s(u)}\mid u\in\phi\}\qquad\text{ and }\qquad
  \cod(\phi)= \{1_{t(u)}\mid u\in \phi\}
\end{equation*}
where $1_x$ denotes the identity arrow on an object $x\in P_0$. The
  domain algebra induced equals the Boolean algebra of subidentities.
The antidomain and anticodomain maps are therefore given again by
relative complementation $\adom(\phi) = \uno \setminus \dom(\phi)$ and
$\acod(\phi) = \uno \setminus \cod(\phi)$ within the domain algebra.
Finally, unfolding definitions shows that
\begin{equation*}
  \fDia{\phi}p = \{1_{s(u)}\mid u\in \phi \land t(u) \in
  p\}\qquad\text{ and }\qquad \fBox{\phi}p = \{1_{s(u)} \mid
  u\in\phi \Rightarrow t(u) \in p\},
\end{equation*}
where $p \subseteq \uno$ is some set of identity
arrows. Reachability along a relation has now been replaced by
  reachability along a set of paths. Similar expressions for backward
modalities can be obtained again by swapping source and target maps in
the right places.

The relational model and the path model are very similar. In fact the
relational model can be obtained from the path model by applying a
suitable homomorphism of modal Kleene algebras.

\subsection{Higher globular Kleene algebras}
\label{SS:ModalKleene}

We now extend the axiomatisations in the previous sections to a new
notion of globular $n$-dimensional modal Kleene algebra. First, we
provide axioms for $n$-dimensional dioids that satisfy lax interchange
laws between multiplications of different dimension, similar to those
of concurrent Kleene algebra~\cite{HoareMollerStruthWehrman11}.  We
then extend it with domain operations of different dimension and add
further axioms that capture globularity. Finally we equip these
algebras with star operations for each dimension and impose novel lax
interchange laws between compositions and stars of different
dimension.

\subsubsection{$n$-Dioid}
\label{SSS:nDioid}
A \emph{$0$-dioid} is a bounded distributive lattice; a
\emph{$1$-dioid} is a dioid. More generally, for $n\geq 1$, an
\emph{$n$-dioid} is a structure
$(S,+,0,\mult_{i},\un_{i})_{0\leq i <n}$ satisfying the following
conditions:
\begin{enumerate}[{\bf i)}]
\item $(S,+,0,\mult_i,\un_i)$ is a dioid for $0\leq i < n$,
\item the following \emph{lax interchange laws} hold for all
  $0\leq i < j < n$:
\begin{equation}
\label{E:WeakExchange}
(x\mult_j x') \mult_i (y \mult_j y')
\leq
(x\mult_i y) \mult_j (x' \mult_i y'),
\end{equation}
\item higher-dimensional units are idempotents of lower-dimensional
  multiplications,  for $0\leq i < j < n$,
\begin{equation}
\label{E:Completeness}
\un_j \mult_i \un_j = \un_j 
\end{equation}

\end{enumerate}
With lax interchange laws we need not worry about an Eckmann-Hilton
collapse.

\subsubsection{Domain $n$-semirings}
\label{SSS:nSemiringWithDomain}
For $n=0$, we stipulate that a \emph{domain $0$-semiring} is a
$0$-dioid. For $n\geq 1$, a \emph{domain $n$-semiring} is an $n$-dioid
$(S,+,0,\mult_i,\un_i)_{0\leq i <n}$ equipped with $n$ domain maps
$\dom_i : S \fl S$, for all $0\leq i <n$, satisfying the following
conditions:
\begin{enumerate}
\item $(S,+,0,\mult_i,\un_i, \dom_i)$ is a domain semiring,
\item $\dom_{i+1}\circ \dom_i = \dom_i$.
\end{enumerate}

For $0\leq i < n$, the set $S_{d_i} = \dom_i(S)$ is called the
\emph{$i$-dimensional domain algebra} and denoted by
$S_i$. Furthermore, to distinguish elements of different dimensions
$0 \leq i < j < n$, we henceforth denote elements of $S_i$ by
$p,q,r, \dots$, elements of $S_j$ by $\phi, \psi, \xi, \dots$, and
other elements of $S$ by $A,B,C, \dots$ This simplifies reading proofs
where elements of different dimension are interacting.  For any
natural number $k$, the \emph{$k$-fold $i$-multiplication} of an
element $A$ of $S$, for $0\leq i < n$, is defined by

\[
A^{0_i} = 1_i,
\qquad
A^{k_i} = A \mult_i A^{(k-1)_i}.
\]

The axioms {\bf ii)} and {\bf iii)} from~\eqref{SSS:nDioid} for
$n$-dioids provide the basic algebraic structure for reasoning about
higher rewriting systems. Indeed, the dependencies between
multiplications of different dimension expressed by the lax
interchange laws capture the lifting of the equational interchange law
for $n$-categories, while the idempotence of $i$-multiplication for
the $j$-unit expresses completeness of the set of $j$-dimensional
cells in an $n$-category with respect to $i$-composition. In this way,
these axioms begin to capture the higher dimensional character of
polygraphs, as is explained in~\eqref{SSS:PolygraphicalModels},
in which we provide a model of this structure based on polygraphs.
The domain axiom {\bf ii)} from~\eqref{SSS:nSemiringWithDomain}
further captures characteristics of dimension, which are expressed
abstractly in the following proposition.

\begin{prop}\label{Prop:nDomainSemirings}
In any domain $n$-semiring $S$ such that $n\ge 1$, for all $0\leq i < j <n$, 
\begin{enumerate}[{\bf i)}]
\item $\dom_j\circ \dom_i = \dom_i$,
\item $\dom_j(\un_i) = \un_i$,
\item $\un_i \leq \un_j$,
\item $S_i \subseteq S_j$,
\item $(S_j,+,0,\mult_{i},\un_{i},d_{i})$ is a domain sub-semiring of $(S,+,0,\mult_{i},\un_{i},d_{i})$ and $\dom_i(S_j) = S_i$,
\item $(S_j , + , 0 , \mult_k, \un_k, \dom_k)_{0\leq k \leq i}$ is a domain sub-$(i+1)$-semiring of $(S,+,0,\mult_{k},\un_{k},d_{k})_{0\leq k \leq i}$, 
\item $(S_j, + , 0 , \mult_j, \un_j)$ is a $0$-dioid.
\end{enumerate} 
\end{prop}
\begin{proof}
  The first identity is proved by a simple induction on axiom
  {\bf{ii)}} in~\eqref{SSS:nSemiringWithDomain}. The second one
  quickly follows, since $\dom_i(\un_i) = \un_i$ follows from the
  domain semiring axioms, and thus
$\dom_j(\un_i) = \un_i$ using {\bf i)}.
The third identity is again a direct consequence, since by {\bf ii)}
we know that $\un_i \in S_j$, and that $\un_j$ is the greatest element of $S_j$.
The fourth one follows since $x \in S_i$ if, and only if,
$\dom_i(x) = x$, which is equivalent to $\dom_j(x) = x$ by {\bf i)}.
The fifth identity is verified by noticing that the inclusion
$S_j \hookrightarrow S$ is a morphism of domain semirings with the
operation $\mult_i$. Furthermore, since $\dom_i(S_j) \subseteq S_i$
and $S_i \subseteq S_j$, we have $\dom_i(S_j) = S_i$. Noticing that, in fact, $S_j \hookrightarrow S$ is
a morphism of domain semirings with the operation $\mult_k$ for any
$0 \leq k \leq i$ gives us {\bf vi)}. The final result follows from basic properties of domain semirings.
\end{proof}

For any $n$-semiring $S$, we denote by $S^{op}$ the $n$-semiring in
which the order of each multiplication operation has been reversed. An
$n$-semiring $S$ is a \emph{codomain $n$-semiring} if $S^{op}$ is a
domain $n$-semiring. The codomain operations are denoted by $\cod_i$.
A \emph{modal $n$-semiring} is an $n$-semiring with domains and
codomains, in which the coherence conditions
$\dom_i\circ \cod_i=\cod_i$ and $\cod_i\circ \dom_i=\dom_i$ hold for
all $0\le i<n$.

\subsubsection{Remarks} 
Section~\eqref{PathModelOne} explains that the path algebra $K(P)$
defined as the power set of $1$-cells in the free category generated
by a $1$-polygraph $P = (P_0, P_1)$ is a model of modal
$1$-semirings. The domain algebra $K(P)_d$ is isomorphic to the power
set of $P_0$. According to~\eqref{SSS:DomainSemiringsOne}, in the
general case of a domain semiring $(S, + , 0, \cdot, 1, d)$, the
domain algebra $S_d$ forms a bounded distributive lattice with $+$ as
join, $\cdot$ as meet, $0$ as bottom and $1$ as top. This is why we
consider a $0$-dioid as a bounded distributive lattice. The
idempotence and commutativity of multiplication reflect the algebraic
properties of a set of identity $1$-cells.

In Section~\ref{SS:ModelsHigherMKA} we construct
higher path algebras over $n$-polygraphs and show that
these form models of modal $n$-semirings. In this
case it makes sense that $(S_i, + , 0 , \mult_i, \un_i)$ is a
$0$-dioid, since an $i$-cell $f : u \fl v$ of an $n$-category $\Cr$ is a
$0$-cell in the hom-category $\Cr(u,v)$.

\subsubsection{Diamond operators}\label{SSS:Diamonds}
Let $S$ be a modal $n$-semiring.  We introduce \emph{forward} and
\emph{backward $i$-diamond} operators defined via (co-)domain
operations in each dimension by analogy
to~\eqref{SSS:modal-semirings}. For any $0\leq i <n$, $A\in S$ and
$\phi \in S_i$, we define
\begin{equation}
\label{E:DiaDefinition}
\fDia{A}_i (\phi) = \dom_i (A \mult_i \phi)
\quad\text{and}\quad
\bDia{A}_i (\phi) = \cod_i  (\phi\mult_i A).
\end{equation}
These diamond operations have all of the properties listed in~\eqref{SSS:ModalitiesKleeneOne} with respect to $i$-multiplication and
elements of $S_i$. As before, antidomains are required to express box
operators.

\subsubsection{$p$-Boolean domain semirings}\label{SSS:pBooleanSemiring}
For $p$ and $n$ such that $0\leq p < n$, a domain $n$-semiring
$(S,+,0,\mult_i,\un_i,d_i)_{0\leq i <n}$ is \emph{$p$-Boolean} if it
is augmented with $(p+1)$ maps
\[
(\adom_i : S \fl S)_{0\leq i \leq p}
\]
such that for all $0\leq i \leq p$, the following conditions are satisfied:
\begin{enumerate}
\item $(S , +, 0 , \mult_i, \un_i, \adom_i)$ is a Boolean domain semiring,
\item $\dom_i = \adom_i^2$.
\end{enumerate}
By definition, a $0$-Boolean domain $1$-semiring is a Boolean domain semiring, and by convention we define a \emph{$0$-Boolean domain $0$-semiring} as a Boolean algebra.

We define a \emph{$p$-Boolean codomain semiring} as an $n$-semiring such that its opposite
$n$-semiring is a $p$-semiring with antidomains. In this case the
anticodomain operations are denoted $\acod_i$.

\begin{rem}
  The key difference between modal $n$-semirings and their $p$-Boolean
  counterparts is that the latter are equipped with negation
  operations in their lower dimensions. Indeed, in a $p$-Boolean modal
  Kleene algebra $K$, for every $0\leq i \leq p$, the tuple
\[
(K_i, + , 0 , \mult_i, \un_i , \adom_i)
\]
is a Boolean algebra. For this reason, we denote the restriction of
$\adom_i$ to $K_i$ by $\neg_i$. Furthermore, as in~\eqref{SSS:ModalitiesKleeneOne}, for $0\leq j \leq p$, $A \in K$ and
$\phi \in K_j$ we can define \emph{forward} (resp.~\emph{backward})
\emph{box operators}
\[
\fBox{A}_j(\phi) := \neg_j(\fDia{A}_j(\neg_j \phi))
\qquad
\text{ and }
\qquad
\bBox{A}_j(\phi) := \neg_j(\bDia{A}_j(\neg_j \phi)).
\]
\end{rem}

 \subsubsection{Globular modal $n$-semiring}
 \label{SSS:GlobularSemiring}

 A modal semiring $S$ is \emph{globular} if the following
 \emph{globular relations} hold for $0\leq i < j <n$ and $A,B \in K$:

\begin{minipage}{.45\textwidth}
\begin{align}
\dom_i \circ \dom_j=\dom_i, \;\; \text{and} \;\; \dom_i \circ \cod_j = \dom_i, \quad\label{GlobularAxiomsIa}\\
\cod_i \circ \dom_j=\cod_i, \;\;\text{and} \;\; \cod_i\circ \cod_j = \cod_i, \quad\label{GlobularAxiomsIb}
\end{align}
\end{minipage}
\begin{minipage}{.5\textwidth}
\begin{align}
\dom_j(A\mult_i B) = \dom_j(A)\mult_i \dom_j(B), \qquad\label{GlobularAxiomsIIa}\\ 
\cod_j(A\mult_i B) = \cod_j(A)\mult_i \cod_j(B). \qquad \label{GlobularAxiomsIIb}
\end{align}
\end{minipage}
\vskip.5cm Any $A\in S$ can be represented diagrammatically with
respect to its $i$- and $j$-borders, for $i<j$:
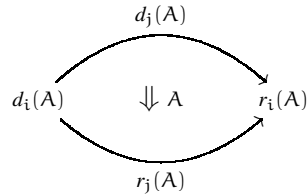
\[
\xymatrix@C=2em@R=1.5em{
{\scriptstyle \dom_i(A)}
	\ar@/^2pc/[rr] ^-{{\scriptstyle\dom_j(A)}}
	\ar@/_2pc/[rr] _-{{\scriptstyle\cod_j(A)}}
&
\Downarrow {\scriptstyle A}
&
{\scriptstyle \cod_i(A)}
}
\]
Intuitively, $A$ is a \emph{collection} of cells and, for
$k \in \{ i, j\}$, $\dom_k (A)$ (resp.~$\cod_k (A)$) is a
\emph{collection} of $k$-cells each of which is the $k$-source
(resp.~$k$-target) of some cell belonging to $A$. In
Section~\ref{SS:ModelsHigherMKA}, this intuition is grounded in the
polygraphic model.

Below are diagrams for $i$- and $j$-multiplication
with respect to $i$- and $j$-borders:
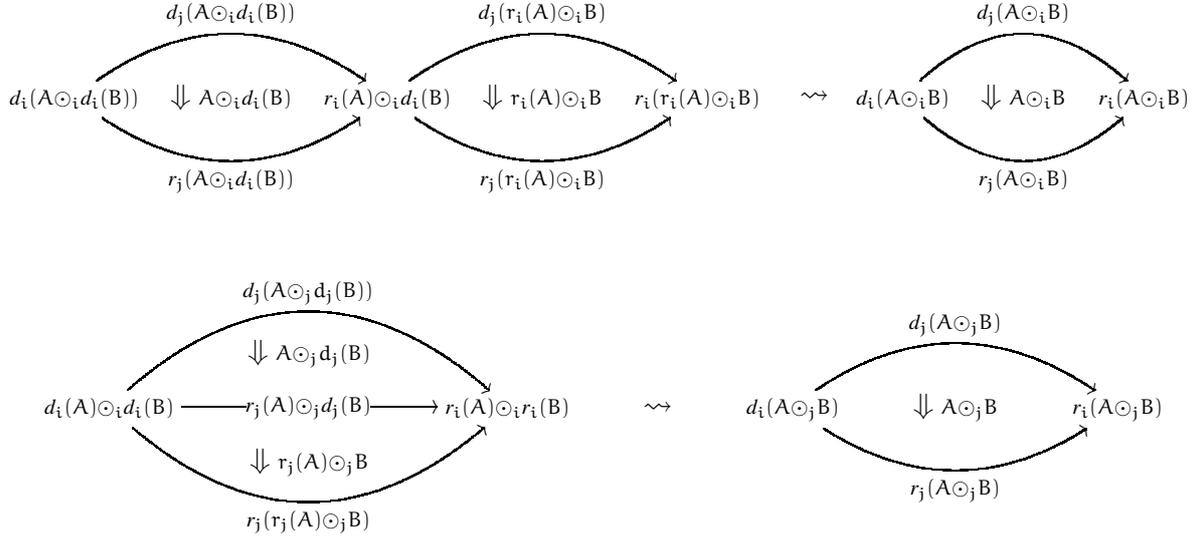
\[
\xymatrix@C=0.25em@R=1.5em{
{\scriptstyle\dom_i(A \mult_i \dom_i(B))}
	\ar@/^2pc/[rr] ^-{\scriptstyle\dom_j(A \mult_i \dom_i(B))}
	\ar@/_2pc/[rr] _-{\scriptstyle\cod_j(A \mult_i \dom_i(B))}
&
 \Downarrow {\scriptstyle A \mult_i \dom_i(B)}
&
{\scriptstyle\cod_i(A) \mult_i \dom_i(B)}
	\ar@/^2pc/[rr] ^-{\scriptstyle\dom_j(r_i(A)\mult_i B)}
	\ar@/_2pc/[rr] _-{\scriptstyle\cod_j(r_i(A)\mult_i B)}
&
\Downarrow {\scriptstyle r_i(A)\mult_i B }
&
{\scriptstyle\cod_i(r_i(A)\mult_i B)}
&
\ar@{}[rr] |(0.2){\rightsquigarrow}
&
&
{\scriptstyle\dom_i(A \mult_i B)}
	\ar@/^2pc/[rr] ^-{\scriptstyle\dom_j(A\mult_i B)}
	\ar@/_2pc/[rr] _-{\scriptstyle\cod_j(A\mult_i B)}
&
\Downarrow  {\scriptstyle A\mult_i B}
&
{\scriptstyle\cod_i(A\mult_i B)}
}
\]

\vskip.4cm

\begin{minipage}[c]{.5\textwidth}
\xymatrix@C=1.6em@R=0.5em{
&
\Downarrow {\scriptstyle A \mult_j d_j(B)}
&
\\
{\scriptstyle\dom_i(A) \mult_i \dom_i(B)}
	\ar@/^3pc/[rr] ^-{\scriptstyle\dom_j(A \mult_j d_j(B))}
	\ar[rr] |{\scriptstyle\cod_j(A)\mult_j\dom_j(B)}
	\ar@/_3pc/[rr] _-{\scriptstyle\cod_j(r_j(A)\mult_j  B)}
&
&
{\scriptstyle\cod_i(A) \mult_i \cod_i(B)}
\\
&
\Downarrow {\scriptstyle r_j(A)\mult_j B}
&
}
\end{minipage}
\quad $\rightsquigarrow$ \quad
\begin{minipage}[c]{.5\textwidth}
\xymatrix@C=1.6em@R=0.5em{
&
&
\\
{\scriptstyle\dom_i(A \mult_j B)}
	\ar@/^2pc/[rr] ^-{\scriptstyle\dom_j(A\mult_j B)}
	\ar@/_2pc/[rr] _-{\scriptstyle\cod_j(A\mult_j B)}
&
\Downarrow {\scriptstyle A\mult_j B}
&
{\scriptstyle\cod_i(A\mult_j B)}
\\
&
&
}
\end{minipage}
\bigskip

These show that multiplication of elements in a Kleene algebra amounts
to to multiplying their restrictions to the appropriate domain or
range as
\[
A \mult_i B = (A \mult_i \cod_i(A)) \mult_i (\dom_i(B) \mult_i B) = (A \mult_i \dom_i(B)) \mult_i (\cod_i(A) \mult_i B),
\]
using properties of domain semirings~\eqref{SSS:DomainSemiringsOne}
and compatibility of these restrictions with globular relations.

\subsubsection{Modal $n$-Kleene algebra}\label{SSS:ModalKleeneAlgebra}
An \emph{$n$-Kleene algebra} is an $n$-dioid $K$ equipped with Kleene
stars $(-)^{\ast_i} : K\rightarrow K$ satisfying
\begin{enumerate}[{\bf i)}]
\item $(K, + , 0 , \mult_i , \un_i, (-)^{\ast_i})$ is a Kleene algebra for $0\leq i < n$,
\item For $0\leq i < j < n$, the Kleene star $(-)^{*_j}$ is a
  \emph{lax morphism} with respect to the $i$-whiskering of
  $j$-dimensional elements on the right (resp.~left). Hence, for all
  $A\in K$ and $\phi \in K_j$,
\begin{equation}
\label{KleenePlusAxioms}
\phi \mult_i A^{*_j} \leq (\phi \mult_i A )^{*_j},
\qquad\text{ and }\qquad
( \text{resp.~}A^{*_j} \mult_i \phi \leq ( A \mult_i \phi )^{*_j} ).
\end{equation}
\end{enumerate}

As in the case of $1$-Kleene algebras in~\eqref{SSS:mka}, the notions
of ($p$-Boolean) $n$-semiring structures with (co)domains are
compatible with those of $n$-Kleene algebra. Hence, a
\emph{$n$-Kleene algebra with domains} (resp.~\emph{codomains}) is a
$n$-Kleene algebra such that the underlying semiring has domains
(resp.~codomains). When the underlying $n$-semiring is modal, this
yields a \emph{modal $n$-Kleene algebra}. If it is $p$-Boolean, we
have a \emph{$p$-Boolean modal $n$-Kleene algebra}. We call it \emph{globular} when the underlying modal $n$-semiring is.

Finally, note that for $n=2$, we recover the standard concurrent
Kleene algebra axioms~\cite{HoareMollerStruthWehrman11}, except that
$\un_0=\un_1$ and commutativity of $\mult_1$ is normally assumed in
this case.

\subsection{A model of higher modal Kleene algebras}
\label{SS:ModelsHigherMKA}

\subsubsection{Polygraphic model}
\label{SSS:PolygraphicalModels}
Let $P$ be an $n$-polygraph and $\Gamma$ a cellular extension of the
free $(n,n-1)$-category $P_n^\top$. In what follows, write
$A, B, C, \dots$ for sets of $(n+1)$-cells and
$\alpha, \beta, \gamma, \dots$ for individual $(n+1)$-cells.  For any
$k$-cell $\alpha$, the elements $s_i(\alpha)$, $t_i(\alpha)$,
$\iota_k^l(\alpha)$ were defined for $0\leq i \leq k \leq l \leq n+1$
in~\eqref{SSS:Notations} and ~\eqref{SSS:IdentitiesWhiskerings}.  When
$k \leq i$, we define $s_i(\alpha) = t_i(\alpha) =
\iota_k^i(\alpha)$. The $i$-composition of a $k$-cell $\alpha$ and an
$l$-cell $\beta$ for $0 \leq i < k \leq l \leq n+1$ was defined
in~\eqref{SSS:HigherCategories}.  For $0\leq k \leq l < n+1$, we
define
\begin{align*}
\alpha \star_i \beta = \begin{cases}
\iota_k^{i+1}(\alpha) \star_i \beta \mbox{ for } k\leq i <l, \\
\iota_k^{i+1}(\alpha) \star_i \iota_l^{i+1}(\beta) \mbox{ for } l \leq i.
\end{cases}
\end{align*}

An $(n+1)$-modal Kleene algebra $K(P,\Gamma)$, the \emph{full
  $(n+1)$-path algebra} over $\tck{P}_n[\Gamma]$ is given by the
following data:
\begin{enumerate}[{\bf i)}]
\item The carrier set of $K(P,\Gamma)$ is the power set
  $\mathcal{P}(P_{n}^\top[\Gamma])$. 

\item For $0\leq i < n+1$, the binary operation $\mult_i$ on $K(P,\Gamma)$
  corresponds to the lifting of the composition operations of
  $P_{n}^\top[\Gamma]$ to the power-set, that is,  for any $A,B \in K(P,\Gamma)$,
\[
A\mult_i B := \{ \alpha \star_i \beta \ | \ \alpha\in A\land \ \beta\in B \land t_i(\alpha) = s_i(\beta)\}.
\]
\item  For $0\leq i < n+1$, the sets
\[
\uno_i \: = \: \{ \iota^{n+1}_i(u) \; |\; u\in \tck{P}_n[\Gamma]_i \},
\]
are the multiplicative units:
$A \mult_i \uno_i \:=\: \uno_i \mult_i A \: = \: A$. Furthermore, when
$i<j$, the inclusion $\uno_i \subseteq \uno_j$ holds. Indeed, in that
case $\iota_i^{n+1} (u) = \iota_j^{n+1}( \iota_i^j(u))$ by uniqueness
of identity cells, and $\iota_i^j(u)\in P_n^\top(\Gamma)_j $ is a
$j$-cell.
\item The addition in $K(P,\Gamma)$ is set union $\cup$. The ordering
  is therefore set inclusion.
\item The $i$-domain and $i$-codomain maps $\dom_i$ and $\cod_i$ are defined by
\[
\dom_i(A) :=  \{ \iota^{n+1}_i (s_i(\alpha)) \ | \ \alpha \in A \},
\qquad\text{ and }\qquad
\cod_i(A) :=  \{ \iota^{n+1}_i(t_i(\alpha)) \ | \ \alpha \in A \}.
\]
These are thus given by lifting the source and target maps of $\tck{P}_n[\Gamma]$ to the power set. The $i$-antidomain and $i$-anticodomain maps are then given by complementation with respect to the set of $i$-cells:
\[
\adom_i(A) := \uno_i \setminus \{ \iota^{n+1}_i (s_i(\alpha)) \ | \ \alpha \in A \},
\qquad\text{ and }\qquad
\acod_i(A) := \uno_i \setminus \{ \iota^{n+1}_i(t_i(\alpha)) \ | \ \alpha \in A \}.
\]
\item The $i$-star is, for $A^{0_i} := \uno_i$ and
  $A^{k_i} := A \mult_i A^{{(k-1)}_i}$,
\[
A^{\ast_i} = \bigcup_{k \in\mathbb{N}} A^{k_i}.
\]
\end{enumerate}

\begin{prop}\label{Prop:Model}
  For any $n$-polygraph $P$ and cellular extension $\Gamma$ of
  $P_n^\top$, $K(P, \Gamma)$ is an $n$-Boolean $(n+1)$-modal Kleene
  algebra.  The set $\rrs{\Gamma}$ of rewriting steps generated by
  $\Gamma$, defined in Remark~\ref{Rem:Contexts}, is represented in
  $n$-Kleene algebra by
\[
\rrs{\Gamma} = \un_{n} \mult_{n-1} \left( \cdots \mult_2 (\un_2 \mult_1(\un_1 \mult_0 \Gamma \mult_0 \un_1) \mult_1 \un_2) \mult_2 \cdots \right) \mult_{n-1} \un_{n}.
\]
Therefore, $\alpha$ is an $(n+1)$-cell of $\tck{P}_n[\Gamma]$ if, and only if, $\alpha \in (\rrs{\Gamma})^{*_n}$.
\end{prop}

\begin{proof}
It is easy to check that, for $0 \leq i < n+1$, the tuple
\[
(\mathcal{P}((\tck{P}_n[\Gamma])_{n+1}) , \cup, \emptyset , \mult_i,  \uno_i , (-)^{\ast_i}, \dom_i, \cod_i)
\] 
is a modal semiring. 
The fact that it is $n$-Boolean is a result of it being a power-set algebra.

Let $A,A',B,B' \in K(P, \Gamma)$ and $0\leq i < j < n+1$. We wish to
check the lax interchange law
\begin{equation}
\label{Eq:InterchangePoly}
(A\mult_j B) \mult_i (A' \mult_j B')
\subseteq
(A\mult_i A') \mult_j (B \mult_i B').
\end{equation}
It holds since, given $(n+1)$-cells
$\alpha \in A, \alpha' \in A', \beta\in B, \beta' \in B'$, if
$(\alpha \star_j \beta) \star_i (\alpha' \star_j \beta')$ is defined,
then as a consequence of the interchange law for $(n+1)$ categories,
we have
$$(\alpha \star_j \beta)\star_i(\alpha' \star_j \beta') = (\alpha \star_i \alpha') \star_j (\beta \star_i \beta') \in (A\mult_i A') \mult_j (B \mult_i B')$$ 
which gives the desired inclusion \eqref{Eq:InterchangePoly}. This situation is illustrated by the diagram

\bigskip

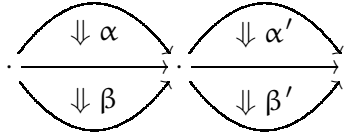
\[
\xymatrix@R=0.4em@C=5em{
\ar@{} [r] |-{\Downarrow \alpha}
&
\ar@{} [r] |-{\Downarrow \alpha'}
&
\\
\cdot
	\ar [r]
	\ar@/^2pc/[r]
	\ar@/_2pc/[r]
&
\cdot
	\ar [r]
	\ar@/^2pc/[r]
	\ar@/_2pc/[r]
&
\cdot
\\
\ar@{} [r] |-{\Downarrow \beta}
&
\ar@{} [r] |-{\Downarrow \beta'}
&
}
\]

\bigskip

\noindent The lax interchange law does not reduce to an equality due
to composition of diagrams of the shape \vskip.25cm
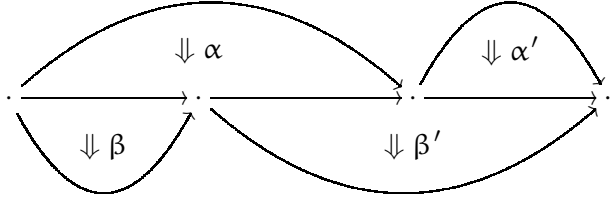
\[
\xymatrix@R=0.4em@C=5em{
&
{\Downarrow \alpha}
&
\ar@{} [r] |-{\Downarrow \alpha'}
&
\\
\cdot
	\ar [r]
	\ar@/^3pc/[rr]
	\ar@/_3pc/[r]
&
\cdot
	\ar [r]
	\ar@/_3pc/[rr]
&
\cdot
	\ar [r]
	\ar@/^3pc/[r]
&
\cdot
\\
\ar@{} [r] |-{\Downarrow \beta}
&
&
{\Downarrow \beta'}
}
\]

\bigskip

\noindent where $\alpha \in A, \alpha' \in A', \beta\in B, \beta' \in
B'$. Indeed, the composition
\begin{equation*}
  (\alpha \star_i \alpha')\star_j(\beta \star_i
  \beta') \in (A\mult_i A') \mult_j (B \mult_i B')
\end{equation*}
is defined, whereas neither $\alpha$ and $\beta$ nor $\alpha'$ and
$\beta'$ are $j$-composable, so that in general the inclusion
\eqref{Eq:InterchangePoly} is strict.

Further, given $0\leq i < j < n+1$, we have
$\uno_j \subseteq \uno_j \mult_i \uno_j$. Indeed, for any $j$-cell
$\alpha$, we have $\alpha \star_i \iota^{n+1}_i(t_i(\alpha)) = \alpha$
because $\iota^{n+1}_i(t_i(\alpha))$ is the $(n+1)$-dimensional
identity cell on the $i$-dimensional target of $\alpha$. Furthermore,
$\iota^{n+1}_i(t_i(\alpha)) \in \uno_i \subseteq \uno_j$, proving the
inclusion. Thus $\uno_j = \uno_j \mult_i \uno_j$ since $(P_n^\top(\Gamma))_j$ is closed under
$i$-composition.

Given $0 \leq i < n$, we have $\dom_{i+1} \circ \dom_i = \dom_i$ since
the $(i+1)$-dimensional border of an identity cell on an $i$-cell $u$
is $u$ itself. Since $\dom_i (\uno_i) = \uno_i$, we equally have
$\dom_{i+1} ( \uno_i) = \uno_i$.

The first two globularity axioms are immediate consequences of the globularity
conditions on the source and target maps of
$P_n^\top(\Gamma)$. Furthermore, for $0\leq i < j < n+1$ and $A,B \in
K(P , \Gamma)$, we have $u \in \dom_j(A \mult_i B)$ if, and only if,
there exist $\alpha \in A$ and $\beta \in B$ such that $u = s_j(\alpha
\star_i \beta) = s_j(\alpha) \star_i s_j(\beta)$, which is equivalent
to $u \in \dom_j(A) \mult_i \dom_j( B)$. Similarly, we show that
$\cod_j(A \star_i B) = \cod_j(A) \mult_i \cod_j(B)$.

Finally, we consider the Kleene star axioms. It is
easy to check that, given a family
$(B_k)_{k \!\! \in  I}$ of elements of $K( P , \Gamma)$ and another element
$A$, we have, for all $0\leq i < n+1$,

\[
A \mult_i \left(\bigcup_{k \in I} B_k \right)= \bigcup_{k \in I} \left(A \mult_i B_k\right)
\quad \text{ and } \quad
\left(\bigcup_{k \in I} B_k \right)\mult_i A = \bigcup_{k \in I} \left( B_k \mult_i A\right).
\]
It then follows by routine calculations that $A^{\ast_i}$ defined
above satisfies, for each $i$, the Kleene star axioms
from~\eqref{SSS:mka}.  It only remains to check that for
$0\leq i < j < n+1$, the $j$-star is a lax morphism for $i$-whiskering
of $j$-dimensional elements on the left (the right case being
symmetric), that is
$\phi \mult_i A^{*_j} \subseteq (\phi \mult_i A )^{*_j}$ for
$\phi \in K(P,\Gamma)_j$ and $A \in K(P,\Gamma)$. By construction,
$K(P,\Gamma)_j$ is in bijective correspondence with
$(P_n^\top(\Gamma))_j$, the set of $j$-cells of
$\tck{P}_n[\Gamma]$. Considering such elements $\phi$ and $A$, we have
$\beta \in \phi \mult_i A^{*_j}$ in the following two cases:
\begin{enumerate}[{\bf i)}]
\item There exist $u\in \phi$ and $\alpha \in A^{+_j}$, where
  $ A^{+_j} := A \mult_j A^{*_j}$ is the Kleene plus operation, such
  that $\beta = u \star_i \alpha$. Since $\alpha \in A^{+_j}$, there
  exist a $k>0$ and cells $\alpha_1, \alpha_2, \dots, \alpha_k \in A$
  such that
\[
\alpha = \alpha_1 \star_j \alpha_2 \star_j \cdots \star_j \alpha_k.
\]
Since $i < j$, the following is a consequence of the
interchange law for $n$-categories:
\begin{align*}
u\star_i (\alpha_1 \star_j \alpha_2 \star_j \cdots \star_j \alpha_k)
=
(u \star_i \alpha_1) \star_j (u \star_i \alpha_2) \star_j \cdots \star_j (u \star_i \alpha_k),
\end{align*}
and thus we have $\beta \in (\phi \mult_i A )^{+_j}$.
\item There exist $u\in \phi$ and $v \in(P_n^\top(\Gamma))_j$ with $v \not\in A$ such that $\beta = u \star_i v$. This is due to the fact that $A^{*_j} = \uno_j + A^{+_j}$. In that case, we have $\beta \in (P_n^\top(\Gamma))_j$, \ie $\beta \in \uno_j$. By the unfold axiom, we have $\uno_j \subseteq (\phi \mult_i A )^{*_j}$, and thus $\beta \in (\phi \mult_i A )^{*_j}$.
\end{enumerate}
The fact that $\alpha$ is an $(n+1)$-cell of $\tck{P}_n[\Gamma]$ if, and only if, $\alpha \in (\rrs{\Gamma})^{*_n}$, follows by definition of $\rrs{\Gamma}$ and the fact that any $(n+1)$-cell of $\tck{P}_n[\Gamma]$ is an $n$-composition of rewriting steps.
\end{proof}

\section{Algebraic coherent confluence}
\label{S:AlgebraicCoherence}

In this section, we present algebraic proofs of the coherent
Church-Rosser theorem and coherent Newman's lemma in 
higher globular Kleene algebras. Apart from the
definitions of these Kleene algebras, these constitute the main
contribution of this article.  First, we revisit abstract rewriting
properties formulated in modal Kleene algebras~\cite{DesharnaisStruth04,Struth02,Struth06} .  We then formalise
notions from higher rewriting needed to prove our results,
introducing \emph{fillers} in the setting of globular modal $n$-Kleene
algebras, which correspond to the notion of fillers for polygraphs
defined in~\eqref{SSS:CoherenceConfluence}.  We also define the
notion of \emph{whiskering} in modal $n$-Kleene algebras, analogous to
the polygraphic definition in~\eqref{SSS:IdentitiesWhiskerings} and describe the properties
thereof needed for our proofs.  The coherent Church-Rosser theorem is
proved in Section~\ref{SS:CoherentChurchRosser}, first in
Proposition~\ref{Th:CRInd} using classical induction and then in
Theorem~\ref{Th:CR} using only the induction axioms for the Kleene
star.  In Section~\ref{SS:CoherentNewman}, we define notions of
termination and well-foundedness in globular modal $p$-Boolean Kleene
algebras and prove Theorem~\ref{Th:Newman}, the coherent Newman's
lemma.

\subsection{Rewriting properties formulated in modal Kleene algebra}\label{SS:RewritingKleeneOne}

Fix a modal Kleene algebra $K$.

\subsubsection{Termination}
\label{SS:CoherentNewmanOne}
An element $x\in K$ \emph{terminates}, or \emph{is Noetherian}~\cite{DesharnaisStruth04}, if
\[
p\leq \fDia{x} p  \Rightarrow p= 0
\]
holds for all $p\in K_\dom$.  The set of Noetherian elements of $K$ is
denoted by $\noeth{K}$. The Galois connections
\eqref{SSS:ModalitiesKleeneOne} lead to the following equivalent
characterisation: $x \in K$ is Noetherian if, and only if, for all $p\in K_\dom$,
\[
\fBox{x} p \leq p \Rightarrow p=1.
\]

\subsubsection{Semi-commutation}
\label{SSS:ConfluenceOne}
Local confluence, confluence and the Church-Rosser property for
abstract rewriting systems are captured in Kleene algebras as follows.
For elements $x,y\in K$, the pair $(x,y)$
\emph{semi-commutes} (resp.~\emph{semi-commutes locally}) if
\[
x^* y^* \leq y^*  x^*
\qquad
(\text{resp. } x y \leq y^*  x^*).
\]
The ordered pair $(x,y)$ semi-commutes \emph{modally}
(resp.~semi-commutes \emph{locally modally}) if
\[
 \fDia{x^*}\circ\fDia{y^*}\leq \fDia{y^*}\circ\fDia{x^*}
\qquad (\text{resp. }\fDia{x}\circ\fDia{y} \leq \fDia{y^*}\circ\fDia{x^*}).
\]
It is obvious that (local) commutation implies (local) modal
commutation; but the converse implication does not hold.  Finally,
$(x,y)$ has the \emph{Church-Rosser property} if
\[
(x + y )^* \leq y^*  x^*.
\]

\subsubsection{Confluence results in Kleene algebras}
\label{SSS:ConfluenceKleeneAlgebras}
The Church-Rosser theorem and Newman's lemma for abstract rewriting
systems are instances of the following specifications in modal Kleene
algebra. In the following subsections we prove higher-dimensional
generalisations of these results.

The Church-Rosser theorem in $K$~{\cite[Thm. 4]{Struth02}} states
that, for any $x,y \in K$, 
\[
x^*y^* \leq y^*x^*
\quad\Leftrightarrow\quad
(x + y )^* \leq y^* x^*.
\]

This does not require modalities. Newman's Lemma in $K$, with $K_\dom$
a complete Boolean algebra~\cite{DesharnaisStruth04}, states that for
any $x,y \in K$ such that $(x+y)\in\noeth{K}$,
\[
\fDia{x}\circ\fDia{y} \leq \fDia{y^*}\circ\fDia{x^*}
\quad\Leftrightarrow\quad
\fDia{x^*}\circ\fDia{y^*}\leq \fDia{y^*}\circ\fDia{x^*}.
\]

Our proofs of the coherent Church-Rosser theorem below are quite
  different from those in Kleene algebra. They require modalities, but
  follow the standard diagrammatic proof quite closely. The parts of
  the proof of coherent Newman's lemma that deal with termination are
  quite different from the standard diagrammatic proof, whereas this
  proof follows the Kleene-algebraic one quite closely. We therefore
  recommend to study this proof~\cite{DesharnaisStruth04} before
  reading the coherent one.
 
\subsection{A coherent Church-Rosser theorem}
\label{SS:CoherentChurchRosser}

Let $K$ be a globular $n$-modal Kleene algebra and $0\leq i < j<n$.
Before defining fillers in globular modal $n$-Kleene algebras, we
first explain the intuition behind the forward diamond operators in
$n$-modal Kleene algebras, defined in~\eqref{SSS:Diamonds}. Given $A \in K$ and $\phi, \phi' \in K_j$,
by definition,
\[
\fDia{A}_j \phi \geq \phi' \iff \dom_j(A \mult_j \phi) \geq \phi'.
\]
In terms of quantification over sets of cells, as for example in the
polygraphic model, this means that \emph{for every} element $u$ in
$\phi'$, \emph{there exist} elements $v$ in $\phi$ and $\alpha$ in $A$
such that the $j$-source (resp.~$j$-target) of $\alpha$ is $u$
(resp.~$v$), as required. This motivates the definitions in the
following paragraph.

\subsubsection{Confluence fillers}
\label{SSS:ConfluenceFillers}
For elements $\phi$ and $\psi$ of $K_j$, we say that an element $A$
in $K$ is a
\begin{enumerate}[{\bf i)}]
\item \emph{local $i$-confluence filler} for $(\phi,\psi)$ if 
\[
\fDia{A}_j(\psi^{*_{i}} \mult_{i} \phi^{*_{i}}) \geq  \phi \mult_{i} \psi,
\]
\item \emph{left} (resp. \emph{right}) \emph{semi-$i$-confluence filler} for $(\phi,\psi)$ if 
\[
\fDia{A}_j(\psi^{*_{i}} \mult_{i} \phi^{*_{i}}) \geq \phi \mult_{i} \psi^{*_i},
\quad \text{(resp. \;\;}
\fDia{A}_j(\psi^{*_{i}} \mult_{i} \phi^{*_{i}}) \geq \phi^{*_i} \mult_{i} \psi
\; \text{)},
\]
\item \emph{$i$-confluence filler} for $(\phi,\psi)$ if 
\[
\fDia{A}_j(\psi^{*_{i}} \mult_{i} \phi^{*_{i}}) \geq \phi^{*_i} \mult_{i} \psi^{*_i},
\]
\item \emph{$i$-Church-Rosser filler} for $(\phi,\psi)$ if 
\[
\fDia{A}_j(\psi^{*_{i}} \mult_{i} \phi^{*_{i}}) \geq (\psi + \phi)^{*_{i}}.
\]
\end{enumerate}

\subsubsection{Remarks}
\label{Rem:GlobularFillers}

In any $n$-Kleene algebra, 
$$(\psi + \phi)^{*_{i}} \geq \phi^{*_i} \mult_{i} \psi^{*_i} \geq \phi
\mult_{i} \psi.$$
This shows that an $i$-Church-Rosser filler for
$(\phi, \psi)$ is an $i$-confluence filler for $(\phi, \psi)$ and that
an $i$-confluence filler for $(\phi, \psi)$ is a local $i$-confluence
filler for $(\phi, \psi)$.

Given an $i$-confluence filler $A$ for $(\phi,\psi)$, the conditions
on domain and codomain in the above definitions imply an
$i$-dimensional globular property of $(\phi,\psi)$.  For all
$p \in K_i$,
\[
  \fDia{\phi^{*_{i}} \mult_{i} \psi^{*_{i}}}_{i}p \leq
  \fDia{\psi^{*_{i}} \mult_{i} \phi^{*_{i}}}_{i} p.
\]

Indeed, writing $A' = A \mult_j (\psi^{*_i} \mult_i \phi^{*_i})$, we have
\begin{align*}
  \fDia{\phi^{*_{i}} \mult_{i} \psi^{*_{i}}}_{i}p
  &= \dom_{i}(\phi^{*_{i}} \mult_{i} \psi^{*_{i}} \mult_{i} p)\\
&\leq \dom_{i}(\dom_{j}(A')\mult_{i} p) \\
&= \dom_{i}(\dom_{j}(A'\mult_{i} p)) \\
&= \dom_{i}(\cod_{j}(A'\mult_{i} p)) \\
&= \dom_{i}(\cod_{j}(A')\mult_{i} p) \\
&\leq \dom_{i}((\psi^{*_{i}} \mult_{i} \phi^{*_{i}})\mult_{i} p)\\
&= \fDia{\psi^{*_{i}} \mult_{i} \phi^{*_{i}}}_{i}p.
\end{align*}

The first step uses the definition of diamonds, the second the fact
that $A$ is an $i$-confluence filler and order-preservation of
$\dom_i$, the third, fourth and fifth the globularity
relations~\eqref{GlobularAxiomsIIa}, \eqref{GlobularAxiomsIa}
and~\eqref{GlobularAxiomsIIb} respectively. The final inequality
follows because $\dom(p\cdot x) = p \cdot \dom(x)$ holds in modal
Kleene algebra, see~\eqref{SSS:DomainSemiringsOne}. For codomains,
opposition implies that
$$\cod_{j}(A') = \cod_j(A \mult_j (\psi^{*_i} \mult_i \phi^{*_i})) = \cod_j(A) \mult_j \cod_j(\psi^{*_i} \mult_i \phi^{*_i}) \leq  \cod_j(\psi^{*_i} \mult_i \phi^{*_i}).$$
The final step is again by definition of diamonds.  Similar results
hold for local and semi-confluence fillers. Thus $\phi$ and $\psi$
commute modally (resp.~locally modally) with respect to
$i$-multiplication.  For this reason, the confluence filler (local
confluence filler) defined in~\eqref{SSS:ConfluenceFillers} can be
represented diagrammatically:

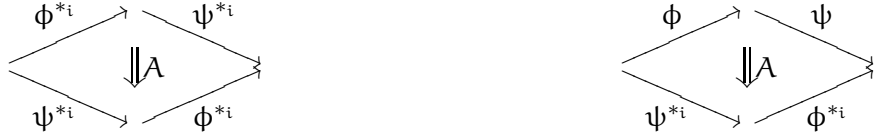
\begin{minipage}{.5\textwidth}
\[
\xymatrix@R=1.5em@C=4em{
&
\ar@{<-}[dl] _-{\phi^{*_{i}}}
\ar[dr] ^-{\psi^{*_{i}}}
\ar@2 "1,2"!<0pt,-15pt>;"3,2"!<0pt,15pt> ^-{A} 
&
\\
\ar[dr] _-{\psi^{*_{i}}}
&
&
\ar@{<-}[dl] ^-{\phi^{*_{i}}}
\\
&&
}
\]
\end{minipage}
\begin{minipage}{.5\textwidth}
\[
\xymatrix@R=1.5em@C=4em{
&
\ar@{<-}[dl] _-{\phi}
\ar[dr] ^-{\psi}
\ar@2 "1,2"!<0pt,-15pt>;"3,2"!<0pt,15pt> ^-{A} 
&
\\
\ar[dr] _-{\psi^{*_{i}}}
&
&
\ar@{<-}[dl] ^-{\phi^{*_{i}}}
\\
&&
}
\]
\end{minipage}

\subsubsection{Whiskers}\label{SSS:Whiskers}
Let $K$ be a globular modal $n$-Kleene algebra. For $0\leq i < j < n$
and $\phi \in K_j$, the \emph{right} (resp.~\emph{left})
\emph{$i$-whiskering} of an element $A \in K$ by $\phi$ is the element
\[
A\mult_i \phi, \qquad \qquad (\text{resp.~} \phi\mult_i A).
\]

In what follows, we list properties of whiskering and define completions.
\begin{enumerate}[{\bf i)}]
\item Let $\phi, \psi \in K_j$ and $A\in K$. We have
\begin{equation}
\label{WhiskeringModalities}
\phi\mult_i \fDia{A}_j(\psi) \leq \fDia{\phi \mult_i A}_j (\phi \mult_i \psi).
\end{equation}
Indeed, since $\phi \mult_j \phi = \phi$, the interchange law gives
\[
\phi \mult_i (A \mult_j \psi) 
= (\phi \mult_j \phi) \mult_i (A \mult_j \psi)
\leq (\phi \mult_i A) \mult_j (\phi \mult_i \psi).
\]
Applying domain on each side yields 
\[
 \phi \mult_i \fDia{A}_j(\psi) = \dom_j (\phi \mult_i (A \mult_j \psi)) \leq \dom_j((\phi \mult_i A) \mult_j (\phi \mult_i \psi)) = \fDia{\phi \mult_i A}_j (\phi \mult_i \psi),
 \]
where we used monotonicity of domains, the definition of diamonds~\eqref{E:DiaDefinition} and one of the globular laws~(\ref{GlobularAxiomsIIa}). A similar argument yields absorption laws for whiskering on the right, as well as corresponding inequalities for backward diamonds.

\item We define \emph{completions} of elements by
  whiskering. Let $A$ be an $i$-confluence filler of a pair
  $(\phi,\psi)$ of elements in $K_j$. The \emph{$j$-dimensional
    $i$-whiskering of $A$} is the following element of $K$:
\begin{equation}
\label{WhiskerCompletion}
(\phi+ \psi)^{*_i}\mult_i A \mult_i (\phi+ \psi)^{*_i}.
\end{equation}
The $j$-star of this element is called the \emph{$i$-whiskered $j$-completion} of $A$. 

\item The $i$-whiskered $j$-completion of a confluence filler
  $A$ absorbs whiskers, that
  is, for each $\xi \leq (\phi + \psi)^{*_i}$
\begin{equation}
\label{WhiskerAbsorption}
\xi \mult_i 	\hat{A}^{*_j}\leq \hat{A}^{*_j} \qquad \text{and} \qquad  \hat{A}^{*_j} \mult_i \xi \leq \hat{A}^{*_j}.
\end{equation}
where $\hat{A}$ is the $j$-dimensional $i$-whiskering of $A$.
Indeed, by definition of $\hat{A}$, 
\begin{align*}
\xi \mult_i \hat{A} \leq \hat{A} \geq \hat{A} \mult_i \xi
\end{align*}
for any $\xi \leq (\phi + \psi)^{*_i}$. Using the fact that $(-)^{*_j}$ is a lax morphism with respect to $i$-whiskering by $j$-dimensional elements, see~\eqref{SSS:mka}, we deduce
\begin{align*}
\xi \mult_i \hat{A}^{*_j} 
\leq (\xi \mult_i \hat{A})^{*_j}
\leq \hat{A}^{*_j},
\end{align*}
where the last inequality holds by monotonicity of $(-)^{*_j}$. A
similar proof shows that
$\hat{A}^{*_j} \mult_i \xi \leq \hat{A}^{*_j}$.
\end{enumerate}

Next we show two proofs of coherent Church-Rosser theorems in globular
$n$-MKA to show the versatility of our approach. The first one
procedes, as usual, by induction on the size of the zig-zag.  It has
no ``closed'' proof in Kleene algebra. The second one uses the
fixpoint induction for the Kleene star. 

\begin{prop}[Coherent Church-Rosser theorem in globular $n$-MKA (by induction)]\label{Th:CRInd}
  Let $K$ be a globular modal $n$-Kleene algebra and $0\leq i <
  j<n$.
  Given $\phi, \psi \in K_j$, an $i$-confluence filler $A$ of
  $(\phi, \psi)$ and any natural number $k\geq 0$, there exists an
  $A_k \leq \hat{A}^{*_j}$ such that
\begin{enumerate}
\item $\cod_j(A_k) \leq   \psi^{*_i} \phi^{*_i}$,
\item $\dom_j(A_k) \geq  (\phi + \psi)^{k_i}$,
\end{enumerate}
where $\hat{A}$ is the $j$-dimensional $i$-whiskering of $A$.
\end{prop}
\begin{proof}
In this proof, juxtaposition of elements denotes $i$-multiplication. 
We reason by induction on $k\geq 0$. 
For $k=0$, we may take $A_0 = \un_i$. Indeed,
\[
\un_i \leq \un_j \leq \hat{A}^{*_j}.
\]
Furthermore, we have $\dom_j(A_0) = \un_i = (\phi + \psi)^{0_i}$ and $\cod_j(A_0) = \un_i \leq \psi^{*_i} \phi^{*_i}$.

For $k> 0$, supposing that $A_{k-1}$ is constructed, we set
\[
A_k = ((\phi+\psi)A_{k-1})\mult_j(A' \phi^{*_i}),
\]
where $A' = A \mult_j (\psi^{*_i} \phi^{*_i})$.
We first show that $\dom_j(A_k) \geq (\phi + \psi)^{k_i}$:

\begin{minipage}{.5\textwidth}
\begin{align*}
\dom_j(A_k) &= \dom_j(((\phi+\psi)A_{k-1})\mult_j(A'\phi^{*_i})) \\
	&= \dom_j(((\phi+\psi)A_{k-1})\mult_j \dom_j(A'\phi^{*_i})) \\
	&= \dom_j(((\phi+\psi)A_{k-1})\mult_j \dom_j(A')\phi^{*_i}) \\
	&\geq \dom_j(((\phi+\psi)A_{k-1})\mult_j \phi^{*_i}\psi^{*_i} \phi^{*_i}) \\
	&=\dom_j((\phi+\psi)A_{k-1}) \\
	&=(\phi+\psi)\dom_j(A_{k-1}) \\
	&=(\phi+\psi)(\phi + \psi)^{(k-1)_i} \\	
	&=(\phi + \psi)^{k_i}.
\end{align*}
\end{minipage}
\begin{minipage}{.5\textwidth}
\[
\xymatrix@R=3em{
\cdot
	\ar[rr] ^-{\scriptstyle(\phi+\psi)}
	\ar[dddrrr] _-{\scriptstyle\psi^{*_i}}
&
&
\cdot
	\ar@{<->}[rrrr] ^-{\scriptstyle(\phi+\psi)^{k-1}}
	\ar[ddrr] |{\scriptstyle\psi^{*_i}}
	\ar@{}[dddr] |-{A'\phi^{*_i}}
&
&
	\ar@{}[dd] |(.3){(\phi+\psi)A_{k-1}}
&
&
\cdot
	\ar@{<-}[ddll] ^-{\scriptstyle\phi^{*_i}}
\\
&&&&&&
\\
& 
& 
&
&
\cdot
	\ar@{<-}[dl] ^-{\scriptstyle\phi^{*_i}}
&
&
\\
&
&
&
\cdot 
&
&
&
}
\]
\end{minipage}
The first step unfolds the definition of $A_k$, the second uses axiom
{\bf{ii)}} from~\eqref{SSS:DomainSemiringsOne} and the third
globularity~\eqref{GlobularAxiomsIIa}. The inequality in the fourth
step is by hypothesis that $A$ is an $i$-confluence filler, and the
fifth is a consequence of the fact that
\[
((\phi+\psi)A_{k-1})\mult_1 (\phi^{*_i}\psi^{*_i}\phi^{*_i}) = (\phi+\psi)A_{k-1},
\]
which in turn holds because
\[
\cod_j((\phi+\psi)A_{k-1}) = (\phi+\psi)\cod_j(A_{k-1}) \leq \phi^{*_i}\psi^{*_i}\phi^{*_i}.
\]
The sixth step is again a consequence of globularity~\eqref{GlobularAxiomsIIa}, the seventh follows from the induction hypothesis, and the last equality is by definition of the $k$-fold $i$-multiplication.

Next we show $\cod_j(A_k) \leq \psi^{*_i} \phi^{*_i}$:
\begin{align*}
\cod_j(A_k) &= \cod_j(((\phi+\psi)A_{k-1})\mult_j(A'\phi^{*_i})) \\
	&= \cod_j(\cod_j((\phi+\psi)A_{k-1})\mult_j(A'\phi^{*_i})) \\
	&\leq \cod_j((\phi+\psi)\psi^{*_i}\phi^{*_i}\mult_j(A'\phi^{*_i})) \\
	&\leq \cod_j((\phi^{*_i}\psi^{*_i}\phi^{*_i})\mult_j(A'\phi^{*_i})) \\
	&= \cod_j(d_j(A'\phi^{*_i})\mult_j(A'\phi^{*_i})) \\
	&= \cod_j(A')\phi^{*_i} \\
	&\leq \psi^{*_i}\phi^{*_i}\phi^{*_i} \\
	&= \psi^{*_i}\phi^{*_i}.
\end{align*}
The first equality holds by definition of $A_k$, the second by axiom
{\bf{ii)}} from~\eqref{SSS:DomainSemiringsOne} (for codomains),
the third by the induction hypothesis, the fourth by
$ \phi \leq \phi^{*_i}$ and $\psi \psi^{*_i} \le \psi^{*_i}$. The
fifth step holds since $A$ is an $i$-confluence filler, the sixth by
the fact that $\dom(x) \cdot x = x$, a consequence of axiom {\bf{i)}}
from~\eqref{SSS:DomainSemiringsOne}. Finally, as explained in~\eqref{Rem:GlobularFillers},
$$\cod_{j}(A') = \cod_j(A \mult_j (\psi^{*_i} \mult_i \phi^{*_i})) = \cod_j(A) \mult_j \cod_j(\psi^{*_i} \mult_i \phi^{*_i}) \leq  \cod_j(\psi^{*_i} \mult_i \phi^{*_i}),$$
which gives step seven since $\psi^{*_i} \mult_i \phi^{*_i} \in
K_j$. The final step is due to
$\phi^{*_i} \mult_i \phi^{*_i} = \phi^{*_i}$, a basic consequence of
the Kleene star axioms.

To conclude, it remains to show that $A_k \leq \hat{A}^{*_j}$. By
whisker absorption, described in~\eqref{WhiskerAbsorption}, and the
fact that $A' \leq A \leq \hat{A}$, 
\[
A'\phi^{*_i} \leq \hat{A}\phi^{*_i} = \hat{A}
\qquad \text{and} \qquad
(\phi + \phi)A_{k-1} \leq (\phi + \psi)\hat{A}^{*_j} \leq \hat{A}^{*_j}.
\]
Thus
$ A_k = ((\phi+\psi)A_{k-1})\mult_j(A\phi^{*_i}) \leq \hat{A}^{*_j} \mult_j
\hat{A}^{*_j} = \hat{A}^{*_j}$.
\end{proof}

We now prove an analogous theorem using the implicit fixpoint
induction of Kleene algebra.
  
\begin{thm}[Coherent Church-Rosser in globular $n$-MKA]
\label{Th:CR}
Let $K$ be a globular $n$-modal Kleene algebra and $0 \leq i < j
<n$. If $\phi, \psi \in K_j$ and is an $i$-confluence filler $A\in K$
of $(\phi,\psi)$, then
\[
\fDia{\hat{A}^{*_j}}_j (\psi^{*_i}\phi^{*_i}) \geq (\phi + \psi)^{*_i},
\]
where $\hat{A}$ is the $j$-dimensional $i$-whiskering of $A$. Thus $\hat{A}^{*_j}$ is an $i$-Church-Rosser filler for $(\phi,\psi)$.
\end{thm}
\begin{proof}
  As in the previous proof, $i$-multiplication is denoted by
  juxtaposition.  Let $\phi, \psi$ be in $K_j$, for $0<j<n$, and $A$
  in $K$ be an $i$-confluence filler of $(\phi,\psi)$, with
  $0\leq i<j$. By the left $i$-star induction axiom in~\eqref{SSS:mka}, 
\[
\un_i + (\phi + \psi) \fDia{\hat{A}^{*_j}}_j (\psi^{*_i}\phi^{*_i})  \leq \fDia{\hat{A}^{*_j}}_j (\psi^{*_i}\phi^{*_i}) 
\; \Rightarrow \;
(\phi + \psi)^{*_i} \leq \fDia{\hat{A}^{*_j}}_j (\psi^{*_i}\phi^{*_i}).
\]
The inequality
$\un_i \leq \psi^{*_i}\phi^{*_i} \leq \fDia{\hat{A}^{*_j}}_j
(\psi^{*_i}\phi^{*_i})$ holds: by the unfold axiom from~\eqref{SSS:mka}, $\un_i \leq \psi^{*_i}$,
$\un_i \leq \phi^{*_i}$,  which yields the first inequality, and
$\un_j \leq \hat{A}^{*_j}$. Using the latter,
$id_{S_{\dom_j}} = \fDia{\un_j}_j \leq \fDia{\hat{A}^{*_j}}_j$, which
yields
$\psi^{*_i}\phi^{*_i} \leq \fDia{\hat{A}^{*_j}}_j
(\psi^{*_i}\phi^{*_i})$.  It then remains to show that
\[
(\phi + \psi) \fDia{\hat{A}^{*_j}}_j (\psi^{*_i}\phi^{*_i})  \leq \fDia{\hat{A}^{*_j}}_j (\psi^{*_i}\phi^{*_i}).
\]
Distributivity allows us to prove this for each summand
separately:
\begin{itemize}
\item In the case of whiskering by $\phi$ on the left,

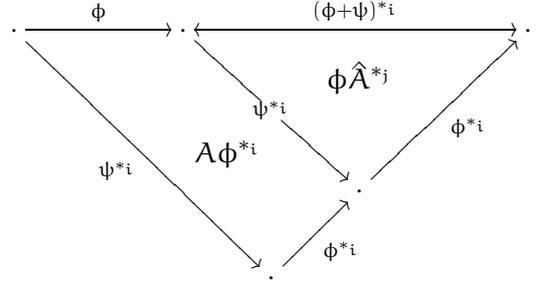
\begin{minipage}{.5\textwidth}
\begin{align*}
\phi \fDia{\hat{A}^{*_j}}_j (\psi^{*_i}\phi^{*_i})
&\leq  \fDia{\phi \hat{A}^{*_j}}_j (\phi \psi^{*_i}\phi^{*_i}) \\
&\leq  \fDia{\phi \hat{A}^{*_j}}_j (\fDia{A}_j( \psi^{*_i}\phi^{*_i})\phi^{*_i}) \\
&\leq \fDia{\phi \hat{A}^{*_j}}_j (\fDia{A\phi^{*_i}}_j( \psi^{*_i}\phi^{*_i}\phi^{*_i})) \\
&\leq \fDia{\phi \hat{A}^{*_j} \mult_j A\phi^{*_i}}_j( \psi^{*_i}\phi^{*_i}) \\
&\leq \fDia{\hat{A}^{*_j} \mult_j \hat{A}}_j( \psi^{*_i}\phi^{*_i}) \\
&\leq \fDia{\hat{A}^{*_j}}_j( \psi^{*_i}\phi^{*_i}). \\
\end{align*}
\end{minipage}
\begin{minipage}{.5\textwidth}
\[
\xymatrix@R=2em{
\cdot
	\ar[rr] ^-{\scriptstyle\phi}
	\ar[dddrrr] _-{\scriptstyle\psi^{*_i}}
&
&
\cdot
	\ar@{<->}[rrrr] ^-{\scriptstyle(\phi+\psi)^{*_i}}
	\ar[ddrr] |{\scriptstyle\psi^{*_i}}
	\ar@{}[dddr] |-{A\phi^{*_i}}
&
&
	\ar@{}[dd] |(.3){\phi \hat{A}^{*_j}}
&
&
\cdot
	\ar@{<-}[ddll] ^-{\scriptstyle\phi^{*_i}}
\\
&&&&&&
\\
& 
& 
&
&
\cdot
	\ar@{<-}[dl] ^-{\scriptstyle\phi^{*_i}}
&
&
\\
&
&
&
\cdot 
&
&
&
}
\]
\end{minipage}
The first step follows from whiskering properties
in~\eqref{SSS:Whiskers}, the second from the hypothesis that $A$ is an
$i$-confluence filler and $\phi\psi^{*_i} \leq \phi^{*_i}
\psi^{*_i}$. The third step is again by whiskering, and the fourth by
definition of diamonds and axiom {\bf{ii)}}
from~\eqref{SSS:DomainSemiringsOne}. The fifth follows by whisker
absorption, \eqref{SSS:Whiskers}, and the last step follows from the
unfold axiom from~\eqref{SSS:mka}, since it implies that
$x \cdot x^* \leq x^*$.

\item In the case of whiskering by $\psi$ on the left,

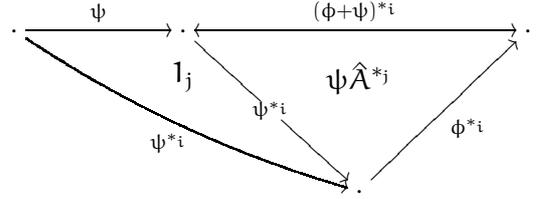
\begin{minipage}{.5\textwidth}
\begin{align*}
\psi \fDia{\hat{A}^{*_j}}_j (\psi^{*_i}\phi^{*_i})
&\leq  \fDia{\psi \hat{A}^{*_j}}_j (\psi \psi^{*_i}\phi^{*_i}) \\
&\leq  \fDia{\psi \hat{A}^{*_j}}_j (\psi^{*_i}\phi^{*_i}) \\
&\leq  \fDia{\hat{A}^{*_j}}_j (\psi^{*_i}\phi^{*_i}). \\
\end{align*}
\end{minipage}
\begin{minipage}{.5\textwidth}
\[
\xymatrix@R=2em{
\cdot
	\ar[rr] ^-{\scriptstyle\psi}
	\ar@/_/[ddrrrr] _-{\scriptstyle\psi^{*_i}}
&
&
\cdot
	\ar@{<->}[rrrr] ^-{\scriptstyle(\phi+\psi)^{*_i}}
	\ar[ddrr] |{\scriptstyle\psi^{*_i}}
	\ar@{}[d] |-{\un_j}
&
&
	\ar@{}[dd] |(.3){\psi \hat{A}^{*_j}}
&
&
\cdot
	\ar@{<-}[ddll] ^-{\scriptstyle\phi^{*_i}}
\\
&&&&&&
\\
& 
& 
&
&
\cdot
&
&
\\
&
&
&
\cdot 
&
&
&
}
\]
\end{minipage}
\end{itemize}
The first step is again by whiskering properties from~\eqref{SSS:Whiskers}, the second by the fact that
$\psi \psi^{*_i} \leq \psi^{*_i}$ which, as explained above, is a
consequence of the unfold axiom from~\eqref{SSS:mka}. Finally, whisker absorption justifies the last
inequality.
\end{proof}

\subsubsection{Remarks}
In Theorem~\ref{Th:CRInd}, the elements $A_k$ satisfy
$\fDia{A_k}_j(\psi^{*_i}\phi^{*_i}) \geq (\phi + \psi)^{k_i}$. This
means that scanning backward along $A_k$ from $\psi^{*_i}\phi^{*_i}$,
we see \emph{at least} all of the "zig-zags" in $\phi$ and $\psi$ of
length $k$, whereas in Theorem~\ref{Th:CR}, the inequality
$\fDia{\hat{A}^{*_j}}_j(\psi^{*_i}\phi^{*_i}) \geq (\phi +
\psi)^{*_i}$ means that scanning back from $\psi^{*_i}\phi^{*_i}$, we
see \emph{at least} all of the zig-zags in $\phi$ and $\psi$ of any
length. However, the elements $A_k$ from Theorem~\ref{Th:CRInd}
satisfy in addition

\[
\bDia{A_k}_j((\phi + \psi)^{k_i}) \leq \psi^{*_i}\phi^{*_i}.
\]
This formulation is consistent with the intuition of paving
\emph{from} zigzags $(\phi + \psi)^{k_i}$ \emph{to} the confluences
$\psi^{*_i}\phi^{*_i}$. Yet this sort of inequality cannot be expected
of the $j$-dimensional $i$-completion of $A$, since in general, using
the path-algebraic intuition, $\hat{A}^{*_j}$ contains additional
cells that go from zigzags to zigzags.  In conclusion, the fact that
the diamonds scan \emph{all} possible future or past states implies
that we must proceed as in Theorem~\ref{Th:CR} when considering
completions, or construct the elements paving precisely what we would
like as in Theorem~\ref{Th:CRInd}.

\begin{cor}\label{Cor:SemiCR}
  Let $K$ be a globular modal $n$-Kleene algebra. If
  $\phi, \psi \in K_j$, for $i<j<n$, for any semi-$i$-confluence
  filler $A\in K$, then
\[
\fDia{\hat{A}^{*_j}}_j (\psi^{*_i}\phi^{*_i}) \geq (\phi + \psi)^{*_i},
\]
where $\hat{A}$ is the $j$-dimensional $i$-whiskering of $A$.
\end{cor}
\begin{proof}
In the case of a left semi-confluence filler, the proof is identical. If $A$ is a right semi-confluence filler, we use the right $i$-star axiom and the proof is given by symmetry.
\end{proof}

\subsection{Newman's lemma in globular modal \texorpdfstring{$n$}{n}-Kleene algebra}
\label{SS:CoherentNewman}

\subsubsection{Termination in \texorpdfstring{$n$}{n}-semirings}

We define the notion of termination, or Noethericity, in a modal
$n$-semiring $K$ as an extension of that for modal Kleene algebras
in~\eqref{SS:CoherentNewmanOne}.  For $0\leq i<j<n$, an element
$\phi\in K_j$ is \emph{$i$-Noetherian} or \emph{$i$-terminating} if
\[
p\leq \fDia{\phi}_i p  \Rightarrow p= 0
\]
holds for all $p\in K_i$.  The set of $i$-Noetherian elements of $K$
is denoted by $\noethi{K}{i}$. When $K$ is a modal $p$-Boolean
semiring, then, as a consequence of the adjunction between diamonds
and boxes outlined in~\eqref{SSS:ModalitiesKleeneOne}, we obtain
an equivalent formulation of Noethericity in terms of the forward box
operator:
\[
  \phi \in\noethi{K }{i} \quad \Leftrightarrow \quad \forall p \in
  K_i, \; \fBox{\phi}_i p \leq p \Rightarrow \un_i \leq p.
\]
Finally, $\phi$ is $i$-\emph{well-founded} if it is $i$-Noetherian in
the opposite $n$-semiring of $K$.

\begin{thm}[Coherent Newman's lemma for globular $p$-Boolean MKA]\label{Th:Newman}
Let $K$ be a globular $k$-Boolean modal Kleene algebra, and $0 \leq i \leq k < j < n$, such that
\begin{enumerate}
\item $(K_i, +, 0, \mult_i, \un_i, \neg_i)$ is a complete Boolean
  algebra,
\item $K_j$ is continuous with respect to $i$-restriction, that is,
  for all $\psi,\psi'\in K_j$ and every family
  $(p_\alpha)_{\alpha\in I}$ of elements of $K_i$, 
\[
\psi \mult_i sup_I (p_\alpha) \mult_i \psi' 
= sup_I (\psi \mult_i p_\alpha \mult_i \psi').
\]
\end{enumerate}
Let $\psi\in K_j$ be $i$-Noetherian and $\phi \in K_j$ $i$-well-founded. If $A$ is a local $i$-confluence filler for $(\phi, \psi)$, then
\[
\fDia{\hat{A}^{*_j}}_j(\psi^{*_i}\phi^{*_i}) \geq \phi^{*_i}\psi^{*_i},
\]
that is, $\hat{A}^{*_j}$ is a confluence filler for $(\phi, \psi)$.
\end{thm}
\begin{proof} We denote $i$-multiplication  by juxtaposition. 
First, we define a predicate expressing restricted $j$-paving. Given $p\in K_i$, let
\[
RP(p) \quad \Leftrightarrow  \quad \fDia{\hat{A}^{*_i}}_j(\psi^{*_i}\phi^{*_i}) \geq \phi^{*_i} p \psi^{*_i}.
\]
By completeness of $K_i$, we may set
$ r := sup\, \{p \; | \; RP(p) \} $.
By continuity of $i$-restriction, we may infer
$RP(r)$. Furthermore, by downward closure of $RP$, 
\[
RP(p) \quad \Leftrightarrow \quad p \leq r.
\]
This, in turn, allows us to deduce
\begin{align*}
\forall p.\ ( RP(\fDia{\phi}_i p ) \wedge RP(\bDia{\psi}_i p )  \Rightarrow RP(p) ) 
&\Leftrightarrow \forall p.\ ( \fDia{\phi}_i p \leq r \wedge \bDia{\psi}_i p \leq r \Rightarrow p\leq r ) \\
&\Leftrightarrow \forall p.\ ( p \leq \bBox{\phi}_i r \wedge  p \leq \fBox{\psi}_i r \Rightarrow p\leq r ) \\
&\Leftrightarrow \bBox{\phi}_i r \leq r \wedge  \fBox{\psi}_i r \leq r.
\end{align*}
It thus suffices to show
$\forall p.\ ( RP(\fDia{\phi}_i p ) \wedge RP(\bDia{\psi}_i p )
\Rightarrow RP(p) )$ in order to conclude that $r = \un_i$, by
Noethericity (resp. well-foundedness) of $\psi$ (resp. $\phi$).

Let $p\in K_i$, set $\fDia{\phi}_i(p) = p_\phi$ and $\bDia{\psi}_i(p)
= p_\psi$ and suppose that $ RP( p_\phi )$ and $ RP(p_\psi)$ hold. Note that
\[
\phi p = \dom_i( \phi p) \phi p = \fDia{\phi}_i(p) \phi p \leq p_\phi \phi, 
\]
since $\dom(x) x = x$ by axiom {\bf{i)}} from~\eqref{SSS:DomainSemiringsOne} and $p \leq \un_i$. We have a
similar inequality for $\psi$, that is $p \psi \leq \psi
p_\psi$. These inequalities, along with the unfold axioms from~\eqref{SSS:mka}, give

\begin{minipage}{.4\textwidth}
\begin{align*}
\phi^{*_i} p \psi^{*_i} 
&\leq \phi^{*_i} p   + \phi^{*_i} \phi p \psi \psi^{*_i} + p \psi^{*_i} \\
&\leq \phi^{*_i} p   + \phi^{*_i} p_\phi \phi  \psi  p_\psi  \psi^{*_i} + p \psi^{*_i}.
\end{align*}
\end{minipage}
\hskip.75cm
\begin{minipage}{.4\textwidth}
\[
\xymatrix@C=.5em{
&& p
\\
&&&
\\
\cdot
	\ar[uurr] ^-{\phi^{*_i} }
&&
}
\xymatrix@C=.5em{
&& 
p 
	\ar[dr] ^-{\psi }
&&
\\
&\ar[ur] ^-{\phi}&&\ar[dr] ^-{\psi^{*_i} }&
\\
\cdot
	\ar[ur] ^-{\phi^{*_i} }
&&
&&\cdot
}
\hskip.5cm
\xymatrix@C=.5em{
p
	\ar[ddrr] ^-{\psi^{*_i} }
&&
\\
&&
\\
&&\cdot
}
\]
\end{minipage}

\noindent The outermost summands are below
$\fDia{\hat{A}^{*_j}}_j(\psi^{*_i}\phi^{*_i})$. Indeed, $id_{S_j} =
\fDia{\un_j}_j \leq \fDia{\hat{A}^{*_j}}_j$ since $\un_j \leq
\hat{A}^{*_j}$, $ p \leq \un_i$ and $\phi^{*_i}, \psi^{*_i} \leq
\psi^{*_i}\phi^{*_i}$.

For the middle summand, we calculate

\begin{minipage}{.3\textwidth}
\begin{align*}
\phi^{*_i} p_{\phi} \phi  \psi  p_{\psi}  \psi^{*_i}
&\leq \phi^{*_i} p_{\phi} \fDia{A}_j(\psi^{*_i} \phi^{*_i})  p_{\psi}  \psi^{*_i} \\
&\leq \fDia{\phi^{*_i} p_{\phi} A p_{\psi}  \psi^{*_i}}_j(\phi^{*_i} p_{\phi} \psi^{*_i} \phi^{*_i} p_{\psi}  \psi^{*_i})\\
&\leq \fDia{\phi^{*_i} p_{\phi} \hat{A} p_{\psi}  \psi^{*_i}}(\fDia{\hat{A}^{*_j}}_j(\psi^{*_i}\phi^{*_i}) \phi^{*_i} p_{\psi}  \psi^{*_i})\\
&\leq \fDia{ \hat{A} }(\fDia{\hat{A}^{*_j} \phi^{*_i}p_{\psi}  \psi^{*_i} }_j (\psi^{*_i} \phi^{*_i} p_{\psi}  \psi^{*_i} ))\\
&\leq \fDia{ \hat{A} \mult_j \hat{A}^{*_j} \phi^{*_i}p_{\psi}  \psi^{*_i} }_j (\psi^{*_i} \phi^{*_i} p_{\psi}  \psi^{*_i} )\\
&\leq \fDia{ \hat{A} \mult_j \hat{A}^{*_j}  }_j (\psi^{*_i} \phi^{*_i} p_{\psi}  \psi^{*_i} ).
\end{align*}
\end{minipage}
\begin{minipage}{.4\textwidth}
\[
\xymatrix@C=1em@R=2em{
&& 
{\scriptstyle p} 
	\ar[dr] ^-{\scriptstyle\psi}
&&
\\
&
\cdot
	\ar[ur] ^-{\scriptstyle\phi}
	\ar[dr] |-{\scriptstyle\psi^{*_i} }
&
\hat{A}
&
\cdot
	\ar[dr] ^-{\scriptstyle\psi^{*_i} }
&
\\
\cdot
	\ar[ur] ^-{\scriptstyle\phi^{*_i} }
	\ar[dr] _-{\scriptstyle\psi^{*_i} }
&
\hat{A}^{*_j}
&
\cdot
	\ar[ur] |-{\scriptstyle\phi^{*_i} }
&&
\cdot
\\
&
\cdot
	\ar[ur] |-{\scriptstyle\phi^{*_i} }
&&&
}
\]
\end{minipage}
\vskip.2cm The first step uses the local $i$-confluence filler
hypothesis, the second whiskering properties from~\eqref{SSS:Whiskers} and the third $RP( p_\phi )$. The fourth
step is again by whiskering properties, and the fifth follows from
axiom {\bf{ii)}} in~\eqref{SSS:DomainSemiringsOne} and the
definition of diamond operators. The final step is by whisker
absorption, see~\eqref{SSS:Whiskers}. A similar arguments yields

\begin{minipage}{.4\textwidth}
\begin{align*}
\fDia{ \hat{A} \mult_j \hat{A}^{*_j}  }_j (\psi^{*_i} \phi^{*_i} p_{\psi}  \psi^{*_i} )
&\leq \fDia{ \hat{A} \mult_j \hat{A}^{*_j}  }_j (\psi^{*_i} \fDia{\hat{A}^{*_j}}_j(\psi^{*_i}\phi^{*_i})) \\
&\leq \fDia{ \hat{A} \mult_j \hat{A}^{*_j}  }_j (\fDia{\psi^{*_i} \hat{A}^{*_j}}_j(\psi^{*_i}\phi^{*_i})) \\
&\leq \fDia{ \hat{A} \mult_j \hat{A}^{*_j} \mult_j \psi^{*_i} \hat{A}^{*_j}}_j(\psi^{*_i}\phi^{*_i}) \\
&\leq \fDia{ \hat{A} \mult_j \hat{A}^{*_j} \mult_j \hat{A}^{*_j}}_j(\psi^{*_i}\phi^{*_i}).\\
\end{align*}
\end{minipage}
\begin{minipage}{.4\textwidth}
\[
\xymatrix@C=1.5em@R=2em{
&& 
{\scriptstyle p} 
	\ar[dr] ^-{\scriptstyle\psi}
&&
\\
&
\cdot
	\ar[ur] ^-{\scriptstyle\phi}
	\ar[dr] ^-{\scriptstyle\psi^{*_i} }
&
\hat{A}
&
\cdot
	\ar[dr] ^-{\scriptstyle\psi^{*_i} }
&
\\
\cdot
	\ar[ur] ^-{\scriptstyle\phi^{*_i} }
	\ar[dr] _-{\scriptstyle\psi^{*_i} }
&
\hat{A}^{*_j}
&
\cdot
	\ar[ur] |-{\scriptstyle\phi^{*_i} }
	\ar@{}[dr] |-{ \hat{A}^{*_j} }
&&
\cdot
\\
&
\cdot
	\ar[ur] |-{\scriptstyle\phi^{*_i} }
	\ar[dr] _-{\scriptstyle\psi^{*_i} }
&&
\cdot
	\ar[ur] _-{\scriptstyle\phi^{*_i} }
&
\\
&&
\cdot
	\ar[ur] _-{\scriptstyle\phi^{*_i} }
&&
}
\]
\end{minipage}
The first step follows from $RP(p_\psi)$ and the second by whiskering
properties. The third step follows from axiom {\bf{ii)}}
in~\eqref{SSS:DomainSemiringsOne} and the definition of diamond
operators as in the preceding calculation. The final step follows from
whisker absorption.  Finally, we observe that
\begin{align*}
\hat{A} \mult_j \hat{A}^{*_j} \mult_j \hat{A}^{*_j} \leq \hat{A}^{*_j},
\end{align*}
and thus by monotonicity of the diamond operator we may conclude that
\[
\phi^{*_i} p_{\phi} \phi  \psi  p_{\psi}  \psi^{*_i}
\leq
\fDia{ \hat{A}^{*_j}}_j (\psi^{*_i} \phi^{*_i}).
\]
This shows that
$\forall p ( RP(p_\phi ) \wedge RP(p_\psi) \Rightarrow RP(p) )$ and
thus that $r= \un_i$, which completes the proof.
\end{proof}

\subsubsection{Remark}
Similarly to the discussion in Remark~\ref{Rem:HDRadvantages} in the
context of polygraphs, the proofs of Theorems~\ref{Th:CR}
and~\ref{Th:Newman} resemble those for $1$-dimensional results modal
Kleene algebra in~\cite{Struth02,DesharnaisStruth04}.  Considering
exclusively the induction axioms and deductions applied to
$j$-dimensional cells yields the same proof structures as for modal
Kleene algebras.  Globular modal $n$-Kleene algebra therefore form a
natural higher-dimensional generalisation of modal Kleene algebras in
which proofs of coherent confluence can be calculated.  The
consistency of the abstract algebraic results from the previous
sections with the point-wise polygraphic results from
Section~\ref{SS:CoherenceConfluence} is made explicit in the next and
final section.

\subsection{Instantiation to polygraphs}\label{SS:ApplicationRewriting}

In the previous sections we have specified and proved Kleene algebraic
versions of Theorems \ref{T:CoherentCRARSFiller} and
\ref{T:NewmanCoherentFiller}. Now we show that they provide faithful
abstractions of the original polygraphic results, instantiating
Theorems \ref{Th:CR} and \ref{Th:Newman} to the polygraphic model of
globular higher Kleene algebras from Section
\ref{SS:ModelsHigherMKA}.

First we add an operation of conversion to higher Kleene algebra, to
capture zig-zag sequences faithfully. We fix an $n$-polygraph $P$ and
a cellular extension $\Gamma$ of the $(n,n-1)$-category $P_n^\top$.

\subsubsection{Converses}\label{SSS:ConverseOne}

A \emph{Kleene algebra with converse}~\cite{Bloom:1995aa} is a Kleene
algebra $K$ equipped with an operation $\conv{(-)} : K \fl K$ that
satisfies

\begin{align*}
\conv{(a+b)} = \conv a + \conv b, \qquad\qquad
\conv{(a\cdot b)} = \conv b \cdot \conv a, \\
\conv{(a^*)} = (\conv a)^*, \qquad
\conv{(\conv a)} = a, \qquad a \leq a \conv{a} a.
\end{align*}
It is an involution that distributes through addition, acts
contravariantly on multiplication and commutes with the Kleene star.
A \emph{modal Kleene algebra with
  converse}~\cite{DBLP:journals/tocl/DesharnaisMS06} is then a modal
Kleene algebra which is also a Kleene algebra with converse.

\subsubsection{$(n,p)$-Kleene algebra}
A modal $(n,p)$-Kleene algebra $K$ is a modal $n$-Kleene algebra
equipped with operations $\con{(-)}{j} : K_{j+1} \fl K_{j+1}$ for
$p \leq j < n-1 $ and an operation $\con{(-)}{n-1} : K \fl K$,
satisfying the axioms listed above for all appropriate
multiplications: for all $\phi,\psi \in K_{j+1}$,
\begin{align*}
\con{(\phi+\psi)}{j} = \con{\phi}{j} + \con{\psi}{j}, \qquad \qquad
\con{(\phi\mult_j \psi)}{j} = \con{\psi}{j} \mult_j \con{\phi}{j} ,\\
\con{(\phi^{*_j})}{j} = (\con{\phi}{j})^{*_j}, \qquad
\con{(\con{\phi}{j})}{j} = \phi, \qquad
\phi \leq \phi \mult_j \con{\phi}{j} \mult_j \phi,
\end{align*}
and $\con{(-)}{n-1}$ satisfies the above axioms with $j=n-1$ and for
any elements of $K$.

Note that for $\phi \in K_i$ with $i < j$, we have
$\con{\phi}{j} = \phi$. This is a consequence of the fact that
$\mult_j$ is idempotent for elements of $K_i$.

\subsubsection{Conversion in the polygraph model}
The modal $(n+1)$-Kleene algebra $K(P, \Gamma)$ generated by $P$ and
$\Gamma$, as defined in~\eqref{SSS:PolygraphicalModels}, is a modal
$(n+1, n-1)$-Kleene algebra. For all $\phi \in K(P, \Gamma)_{n}$ and
$A \in K$, 
\begin{align*}
\con{\phi}{n-1} := \{ \ u^- \; | \; u \in \phi \} \qquad \text{and} \qquad \con{A}{n} := \{ \ \alpha^- \; | \; \alpha \in A \}
\end{align*} 
is well defined in the following sense: Every
$\phi \in K(P, \Gamma)_{n}$ is a set of cells of dimension less than
or equal to $n$. Any cell $v$ of dimension $i < n$ is its own
$n$-inverse, since we consider it as an identity. For any $n$-cell
$u$, we know that $u^-$ is well defined since if $u \in P_n^\top$ then
$u^- \in P_n^\top$. The case of $\con{(-)}{n}$ is similar.

\subsubsection{$\Gamma$-coherence properties as fillers}

Recall that $\Gamma$ and $P_n^*$ are themselves elements of $K(P,\Gamma)$, and that in Proposition~\ref{Prop:Model} we observed that
\[
\rrs{\Gamma} = \un_{n} \mult_{n-1}( \cdots \mult_2 (\un_2 \mult_1(\un_1 \mult_0 \Gamma \mult_0 \un_1) \mult_1 \un_2) \mult_2 \cdots) \mult_{n-1} \un_{n},
\]
where $\rrs{\Gamma}$ is the set of cells of $\Gamma$ in context. In
the following, we write $\rrs{P_n}$ for the set of rewriting steps
generated by $P_n$, which can be expressed in $K(P, \Gamma)$ as
\[
\rrs{P_n} = \left( \un_{n-1} \mult_{n-2}( \cdots \mult_2 (\un_2 \mult_1(\un_1 \mult_0 P_n \mult_0 \un_1) \mult_1 \un_2) \mult_2 \cdots) \mult_{n-2} \un_{n-1} \right).
\]
The construction of $K(P,\Gamma)$ is compatible with $\Gamma$-coherence properties in the following sense:

\begin{prop}\label{Prop:GammaCoherence}
  With $\Gamma' := (\rrs{\Gamma})^{*_n}$, 
\begin{enumerate}
\item $\Gamma$ is a (local) confluence filler for $P$ if, and only if, $\Gamma'$ is a (local) $(n-1)$-confluence filler for~$(\con{(\rrs{P_n})}{n-1}, \rrs{P_n})$,
\item $\Gamma$ is a Church-Rosser filler for $P$ if, and only if, $\Gamma'$ is an $(n-1)$-Church-Rosser filler for~$(\con{(\rrs{P_n})}{n-1}, \rrs{P_n})$.
\end{enumerate}
\end{prop}
\begin{proof}
We prove the equivalence in the case of (global) confluence.

Suppose  $\Gamma$ is a confluence filler for $P$. An element
$f^- \star_{n-1} g \in \con{(\rrs{P_n})}{n-1}\mult_{n-1} \rrs{P_n}$
corresponds to a branching $(f,g)$. By hypothesis, there exists an
$\alpha \in \tck{P}_n[\Gamma]$ such that
$s_{n}(\alpha) = f^- \star_{n-1} g$ and $\alpha$ is an $n$-composition
of rewriting steps, so $\alpha \in \Gamma'$. Furthermore, the
$n$-target of $\alpha$ is a confluence, so
$\alpha \in \Gamma' \mult_n (\rrs{P_n} \mult_{n-1}
\con{(\rrs{P_n})}{n-1} )$. In terms of equations, this means that
\[
\con{(\rrs{P_n})}{n-1}\mult_{n-1} \rrs{P_n} 
\subseteq \dom_n \left( \Gamma' \mult_n (\rrs{P_n} \mult_{n-1} \con{(\rrs{P_n})}{n-1} ) \right)
= \fDia{\Gamma'}_n \left( \rrs{P_n} \mult_{n-1} \con{(\rrs{P_n})}{n-1} \right),
\]
that is, $\Gamma'$ is an $(n-1)$-confluence filler for $\left(\con{(\rrs{P_n})}{n-1}, \rrs{P_n}\right)$.

Conversely, if $\Gamma'$ is an $(n-1)$-confluence filler for
$\left(\con{(\rrs{P_n})}{n-1}, \rrs{P_n}\right)$, then, for any
branching $(f,g)$, we know that
$f^- \star_{n-1} g \in \dom_i{(\Gamma' \mult_n (\rrs{P_n} \mult_{n-1}
  \con{(\rrs{P_n})}{n-1} ))}$. Thus there is a cell
$\alpha \in \Gamma'$ with $n$-source $f^- \star_{n-1} g$ and whose
$n$-target is a confluence. Since $\alpha \in \Gamma'$, it must be a
composition of rewriting steps of $\Gamma$. Thus $\Gamma$ is a
confluence filler for $P$, and the remaining cases are similar.
\end{proof}

Proposition~\ref{Prop:GammaCoherence} allows us to instantiate our
main results, Theorems \ref{Th:CR} and \ref{Th:Newman}, in the
polygraphic model and obtain the original theorems of polygraphic
rewriting as corollaries. Theorems \ref{T:CRPoly} and~\ref{T:NMPoly}
below correspond exactly to Theorems \ref{T:CoherentCRARSFiller}
and~\ref{T:NewmanCoherentFiller}, but are obtained through Kleene
algebraic proofs.

\begin{thm}[Church Rosser for $n$-polygraphs]\label{T:CRPoly}
Let $P$ be an $n$-polygraph and $\Gamma$ a cellular extension of $P_n^\top$. Then $\Gamma$ is a confluence filler for $P$ if, and only if, $\Gamma$ is a Church-Rosser filler for $P$.
\end{thm}
\begin{proof}
  Suppose first that $\Gamma$ is a confluence filler for $P$. Using
  the result and notations from Proposition~\ref{Prop:GammaCoherence},
  we know that $\Gamma'$ is an $(n-1)$-confluence filler for
  $(\con{(\rrs{P_n})}{n-1}, \rrs{P_n})$. Applying Theorem~\ref{Th:CR}
  to $K(P,\Gamma)$ for $i=n-1$ and $j=n$ shows
  $\widehat{\Gamma'}^{*_n}$ is an $(n-1)$-Church-Rosser filler for
  $(\con{(\rrs{P_n})}{n-1}, \rrs{P_n})$. Then, using
  $(\rrs{P_n} + \con{(\rrs{P_n})}{n-1})^{*_{n-1}} = P_n^\top$ yields
\begin{align*}
\widehat{\Gamma'}^{*_n} = \left( P_n^\top \mult_{n-1} ( \rrs{\Gamma} )^{*_n} \mult_{n-1} P_n^\top \right)^{*_n} 
				\subseteq \left( (P_n^\top \mult_{n-1} \rrs{\Gamma} \mult_{n-1} P_n^\top)^{*_n} \right)^{*_n} 
				= \Gamma',
\end{align*}
where the first step is by definition, the second uses the fact that
the $n$-star is a lax morphism for $(n-1)$-multiplication, see~\eqref{SSS:ModalKleeneAlgebra}, and the third uses the fact that
$\rrs{\Gamma}$ absorbs whiskers and that $(A^{*_n})^{*_n} =
A^{*_n}$. Since, additionally,
$\Gamma' \subseteq \widehat{\Gamma'}^{*_n}$, $\Gamma'$ is an
$(n-1)$-Church-Rosser filler for $(\con{(\rrs P_n)}{n-1}, \rrs
P_n)$. By Proposition~\ref{Prop:GammaCoherence}, this allows us to
conclude that $\Gamma$ is a Church-Rosser filler for $P$.

For the trivial direction, suppose $\Gamma$ is a Church-Rosser filler
for $P$. Proposition~\ref{Prop:GammaCoherence} implies that $\Gamma'$
is an $(n-1)$-Church-Rosser filler for
$(\con{(\rrs{P_n})}{n-1}, \rrs{P_n})$. As pointed out at the end
of~\eqref{SSS:ConfluenceFillers}, this means that $\Gamma'$ is an
$i$-confluence filler for $(\con{(\rrs{P_n})}{n-1}, \rrs{P_n})$, and
it follows that $\Gamma$ is a confluence filler for $P$.
\end{proof}

\begin{thm}[Newman for $n$-polygraphs]\label{T:NMPoly}
Let $P$ be a terminating $n$-polygraph and $\Gamma$ a cellular extension of $P_n^\top$. Then $\Gamma$ is a local confluence filler for $P$ if, and only if, $\Gamma$ is a confluence filler for $P$.
\end{thm}
\begin{proof}
  Suppose $\Gamma$ is a local confluence filler for $P$. Using the
  result and notations from Proposition~\ref{Prop:GammaCoherence}, we
  know that $\Gamma'$ is an $(n-1)$-local confluence filler for
  $(\con{(\rrs{P_n})}{n-1}, \rrs{P_n})$. We apply
  Theorem~\ref{Th:Newman} to $K(P,\Gamma)$ for $i=n-1$ and $j=n$,
  obtaining that $\widehat{\Gamma'}^{*_n}$ is an $(n-1)$-confluence
  filler for $(\con{(\rrs{P_n})}{n-1}, \rrs{P_n})$. As in the proof of
  the previous theorem, $\widehat{\Gamma'}^{*_n} = \Gamma'$, which
  allows us to conclude that $\Gamma$ is a confluence filler for $P$,
  again by Proposition~\ref{Prop:GammaCoherence}.

  For the trivial direction, suppose  $\Gamma$ is a confluence
  filler for $P$. As above, we deduce that $\Gamma'$ is an
  $(n-1)$-Church-Rosser filler for
  $(\con{(\rrs{P_n})}{n-1}, \rrs{P_n})$. Again, as pointed out at in~\eqref{SSS:ConfluenceFillers}, this means that $\Gamma'$ is a
  local $i$-confluence filler for
  $(\con{(\rrs{P_n})}{n-1}, \rrs{P_n})$, by which we conclude that
  $\Gamma$ is a local confluence filler for $P$ via
  Proposition~\ref{Prop:GammaCoherence}.
\end{proof}

\bigskip

\subsubsection*{Acknowledgements} We wish to thank the anonymous referees for fruitful comments about this article. The fourth author has been supported by the LABEX MILYON (ANR-10-LABX-0070) of Université de Lyon, within the program \emph{Investissements d’Avenir} (ANR-11-IDEX-0007) operated by the French National Research Agency (ANR), and a fellowship at Collegium de Lyon.

\bibliographystyle{alphaurl}
\bibliography{biblioCURRENT}

\end{document}